\documentclass[12pt]{article}
\date{\vspace{-5ex}}
\usepackage{graphicx}
\usepackage{titlesec}
\titleformat*{\section}{\large\bfseries}
\usepackage{amssymb}
\usepackage{marginnote}
\usepackage{subfig}
\usepackage{outlines}
\usepackage{commath}
\usepackage[font=footnotesize]{caption}
\usepackage[textwidth = 7in]{geometry}
\usepackage{mathtools}
\usepackage[dvipsnames]{xcolor}
\stepcounter{secnumdepth}
\stepcounter{tocdepth}

\pagebreak
\usepackage{amsmath}
\DeclareMathOperator{\arcsec}{arcsec}

\newcommand{\e}{\mathrm{e}}

\title{Kagome network with vertex coupling \\ of a preferred orientation}

\author{Marzieh Baradaran$^{1,2}$ and Pavel Exner$^{3,4}$}

\date{\scriptsize 1) Department of Physics, Faculty of Science, University of Hradec Kr\'alov\'e, Rokitansk\'eho 62, 500 03 Hradec Kr\'alov\'e, Czechia \\
2) Department of Mathematics, Faculty of Nuclear Sciences and Physical Engineering, Czech Technical University, B\v rehov\'a 7, 11519 Prague, Czechia \\
3) Doppler Institute for Mathematical Physics and Applied Mathematics, Czech Technical University, B\v rehov\'a 7, 11519 Prague, Czechia \\
4) Department of Theoretical Physics, Nuclear Physics Institute, Czech Academy of Sciences, 25068 \v{R}e\v{z} near Prague, Czechia \\
\emph{marzie.baradaran@yahoo.com, exner@ujf.cas.cz}}

\begin{document}

\captionsetup[figure]{labelfont={bf},labelformat={default},labelsep=period,name={Fig.}}
\maketitle

\begin{abstract}
We investigate spectral properties of periodic quantum graphs in the form of a kagome or a triangular lattice in the situation when the condition matching the wave functions at the lattice vertices is chosen of a particular form violating the time-reversal invariance. The positive spectrum consists of infinite number of bands, some of which may be flat; the negative one has at most three and two bands, respectively. The kagome lattice example shows that even in graphs with such an uncommon vertex coupling spectral universality may hold: if its edges are incommensurate, the probability that a randomly chosen positive number is contained in the spectrum is $\approx 0.639$.
\end{abstract}

%%%%%%%%%%%%%%%%%%%%%%%%
\section{Introduction}
\setcounter{equation}{0}

Quantum graphs proved to be a useful tool to model quantum transport in periodically structured environments, both natural and artificially prepared; the latter are gaining importance in connection with the progress in metamaterial physics. The fact that the way in which the wave functions are coupled in the nodes of the network is vital for the band structure is quite old; one can trace it back the celebrated paper of Kronig and Penney \cite{KP31}. As long as the only requirement one imposes is the conservation of the probability current, mathematically expressed as the self-adjointness of the corresponding Hamiltonian, there is a number of ways how to do that \cite{BK13, Ha00, KS99}: for a vertex $v$ in which $N$ edges meet, the self-adjointness is ensured provided the the boundary-value vectors of the wave functions and their derivatives are matched through the condition
 % ------------- %
\begin{equation}\label{genbc}
 (U-I)\psi(v)+i\ell(U+I)\psi'(v)=0,
\end{equation}
 % ------------- %
where $\ell>0$ is the parameter fixing the length scale and $U$ is an $N\times N$ unitary matrix.

Given such a multitude, one naturally asks about the meaning of the couplings parametrized by different matrices $U$. The simplest class, the so-called $\delta$ coupling with the wave functions continuous at the vertex and the condition \eqref{genbc} being reduced to $\sum_{j=1}^N \psi_j'(v)=\alpha\psi(v)$ with a real parameter $\alpha$, can be understood easily as modeling a potential sharply localized around the vertex \cite{Ex96}. The general coupling \eqref{genbc} can also be interpreted in terms of properly scaled potentials, however, the approximation is considerably more complicated \cite{CET10} and the result has mostly the existence meaning.

A pragmatic approach is to choose the coupling that suits the model in question. Recently a class of couplings unnoticed so far attracted attention, with the motivation coming from an attempt to model the anomalous Hall effect using a lattice graph \cite{SK15}. The said model used the $\delta$ coupling at the lattice nodes which forced the authors to impose by hand a preferential direction. That was a flaw since such an assumption cannot justified on the lattice edges, but it inspired the observation that the family specified by the condition \eqref{genbc} includes vertex couplings that may not be invariant with respect to the time reversal \cite{ET18}. The simplest among them corresponds to the matrix $U$ of the circulant type, with the entries equal to one at the first side diagonal and in the opposite corner, and zero otherwise. A rotational motion associated with such a matrix becomes obvious if we realize that $U$ is nothing but the on-shell scattering matrix of vertex at the momentum $k=\ell^{-1}$.

It appeared that such a coupling has a remarkable topological property, namely that the transport properties of the vertex at \emph{high energies} depend on the vertex \emph{parity}; in this asymptotic regime the vertex remain transparent if the parity is even, while for the odd one we get an effective decoupling of the edges. This effect was illustrated in \cite{ET18} through comparison of band spectra two lattices, the square and the hexagonal one. The spectrum appeared to be dominated by the bands and gaps, respectively, in the sense that the probability that a randomly chosen positive energy belong to the spectrum, as defined by Band and Berkolaiko \cite{BB13}, equals one and zero, respectively.

However, things may not be that simple. In \cite{BET20, BET21}, we investigated another graph with the described vertex coupling, a periodic chain of rings connected either tightly, or loosely through connecting links. The loosely connected chain with vertices of degree three appeared to be effectively decoupled at high energies, but for the tightly connected one the probability of being in the spectrum might or might not equal to one; this happened if the chain had (vertically) the mirror symmetry, otherwise the said probability equaled one half. We also found that despite these differences the spectrum of the loose chain converges to that of the tight one as the lengths of the connecting links shrink to zero, but the convergence was rather non-uniform.

The aim of the present paper is to analyze another class of quantum graphs with the indicated vertex coupling violating the time-reversal invariance, namely lattices of \emph{kagome} and \emph{triangular type}, where the latter can be regarded as the degenerate case of the former. As in other periodic graphs we find that under appropriate rationality conditions such a system exhibit flat bands. Apart from them, there is no effective decoupling here since the vertex parities are always even in the present situation. The probability that a positive energy belongs to a spectral band is equal to two thirds if the kagome lattice is equilateral or degenerate to the triangular one. If the lattice is asymmetric, the probability is different, however, if the edge lengths are \emph{incommensurate}, it takes a fixed value $\approx 0.639$ showing that the universality result derived in \cite{BB13} for Kirchhoff graphs may be valid for a much wider class of vertex couplings.

Let us mention briefly the contents of the paper. In the next section we collect the needed information about the vertex coupling. The spectral problem for the kagome and triangular lattices are solved respectively in Secs.~3 and 4. We derive the appropriate spectral conditions and solve then separately for the positive and negative part of the spectrum.

%%%%%%%%%%%%%%%%%%%%%%%%
\section{Preliminaries}
\setcounter{equation}{0}

Let us describe now the basic setting in more technical terms. We suppose that the motion on the graph edges is free away from the vertices, so that the Hamiltonian acts there as $-\frac{\mathrm{d}^2}{\mathrm{d}x^2}$. Consider a vertex $v$ of degree $N$. Writing the coupling condition \eqref{genbc} with the circulant matrix $U$ described above in components, we get
% -------------- %
\begin{equation}\label{coupling}
(\psi_{j+1}-\psi_{j})+i\ell(\psi_{j+1}^{\prime}+\psi_{j}^{\prime})=0,
\end{equation}
% -------------- %
where $\psi_j,\, j=1,\dots,N,$ are the components of $\psi(v)$ and similarly for $\psi'(v)$. The corresponding on-shell scattering matrix, $S(k) = \frac{k\ell-1 +(k\ell+1)U}{k\ell+1 +(k\ell-1)U}$, can be also expressed in components \cite{ET18} being
 % -------------- %
\begin{equation}\label{sij,onshell}
S_{ij}(k) =\frac{1-\eta^2}{1-\eta^N}\bigg\{ -\eta \, \frac{1-\eta^{N-2}}{1-\eta^2}\,\delta_{ij}+(1-\delta_{ij})\,\eta^{(j-i-1)(\textstyle{\rm mod}\;N)}          \bigg\},
\end{equation}
 % -------------- %
where $\eta:=\tfrac{1-k\ell}{1+k\ell}$. Inspecting the behavior of this expression in the high-energy limit, $\eta\to 1-$, we find that $\lim_{k\to\infty} S(k)=I$ if $N$ is odd, while for $N$ even the limit is different from the unit matrix describing the full separation of the edges. The root of this difference is the fact that $-1$ is an eigenvalue of $U$ if and only if $N$ is even.

The peculiar feature of quantum graphs is that they have a single propagating mode. As a consequence, the motion is free away from the vertex and the scattering matrix makes sense irrespective of the edge lengths. If the edges are semi-infinite, the Hamiltonian of such a star graph has a nonempty discrete spectrum for any $N\ge 3$ and the eigenvalues are
% -------------- %
\begin{equation}\label{Neg,Eig,StarG}
E=-\tan^{2}\frac{m\pi}{N},
\end{equation}
% -------------- %
with $m$ running through $ 1,\cdots,[\tfrac{N}{2}] $ for odd $N$ and $1,\cdots,[\tfrac{N-1}{2}]$ for even $N$.

%%%%%%%%%%%%%%%%%%%%%%%%%
\section{Kagome lattice}
\label{Kag,noneq,section}
\setcounter{equation}{0}

\subsection{The spectral condition}

\begin{figure}[h]
\centering
\includegraphics[scale=0.9]{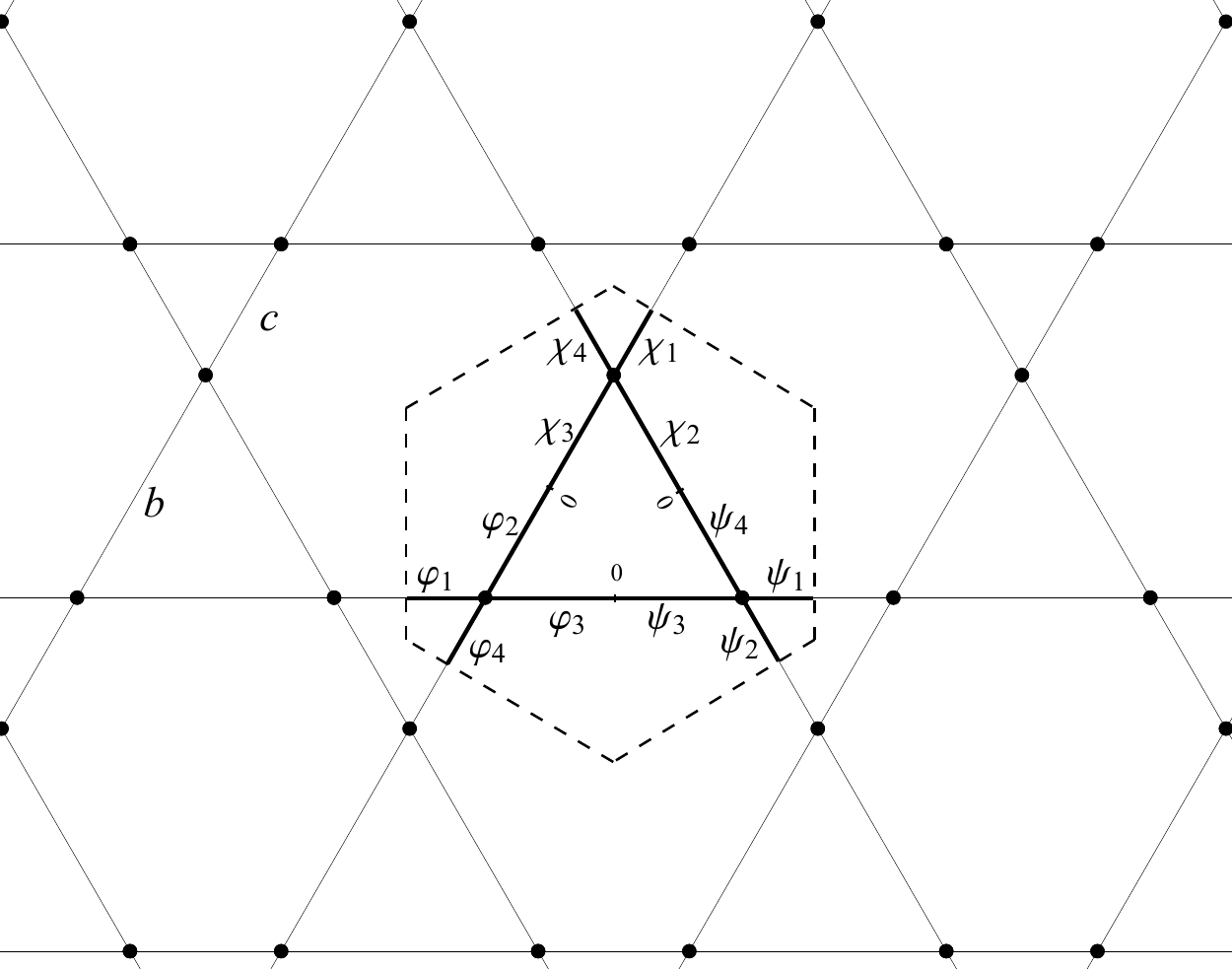}
\caption{ An elementary cell of the kagome network.}
\label{KagUnitCell}
\end{figure}

Our main topic in this paper is spectral properties of periodic quantum graphs of \emph{kagome} type sketched in Fig.~\ref{KagUnitCell}. We suppose that the edge lengths $b$ and $c$ are both positive, postponing the degenerate case of a triangular lattice to the next section, and assume that their sum is fixed, $b+c=d$.
 % -------------- %
 % -------------- %
Since the system is periodic, its spectral analysis relies on the Floquet-Bloch decomposition \cite[Chap.~4]{BK13} which makes it possible to reduce the task to investigation of an elementary cell of the graph which contains three vertices of degree four. Choosing the coordinates on the edges to increase from the left to right, we employ the following Ansatz for the wave function components
 % -------------- %
\begin{align}\label{KagAnsatz}
& \psi_{j}(x)=B_{j}^{+}\e^{ikx}+B_{j}^{-}\e^{-ikx},\qquad x\in[0,\textstyle{\frac12}c], \;\;\qquad j=1,2,\nonumber \\
& \psi_{j}(x)=B_{j}^{+}\e^{ikx}+B_{j}^{-}\e^{-ikx},\qquad x\in[0,\textstyle{\frac12}b], \;\;\qquad j=3,4,\nonumber\\[5pt]
& \varphi_{j}(x)=C_{j}^{+}\e^{ikx}+C_{j}^{-}\e^{-ikx},\qquad x\in[-\textstyle{\frac12}b,0], \qquad j=2,3,\nonumber\\
& \varphi_{j}(x)=C_{j}^{+}\e^{ikx}+C_{j}^{-}\e^{-ikx},\qquad x\in[-\textstyle{\frac12}c,0], \qquad j=1,4,\nonumber \\[5pt]
& \chi_{1}(x)=D_{1}^{+}\e^{ikx}+D_{1}^{-}\e^{-ikx},\qquad x\in[0,\textstyle{\frac12}c],\\
& \chi_{2}(x)=D_{2}^{+}\e^{ikx}+D_{2}^{-}\e^{-ikx},\qquad x\in[-\textstyle{\frac12}b,0],\nonumber\\
& \chi_{3}(x)=D_{3}^{+}\e^{ikx}+D_{3}^{-}\e^{-ikx},\qquad x\in[0,\textstyle{\frac12}b],\nonumber\\
& \chi_{4}(x)=D_{4}^{+}\e^{ikx}+D_{4}^{-}\e^{-ikx},\qquad x\in[-\textstyle{\frac12}c,0]. \nonumber
\end{align}
 % -------------- %
The network is periodic in two independent directions, for the sake of definiteness we assume that they are associated with the unit vectors $(1,0)$ and $\frac12(1,\sqrt{3})$; the periodicity in the remaining direction is a superposition of those two. Consequently, the Floquet conditions at the free ends of the graph cell are
 % -------------- %
\begin{align}\label{KagFloq}
\chi _1 (\textstyle{\frac12}c )&=\e^{i \theta_{1} } \varphi _{4} (-\textstyle{\frac12}c ),\; & \chi _1' (\textstyle{\frac12}c )&=\e^{i \theta_{1} } \varphi _{4}' (-\textstyle{\frac12}c ),
\nonumber \\
\psi _1 (\textstyle{\frac12}c )&=\e^{i \theta_{2} } \varphi _{1} (-\textstyle{\frac12}c ),\; & \psi _1' (\textstyle{\frac12}c )&=\e^{i \theta_{2} } \varphi _{1}' (-\textstyle{\frac12}c ),
\nonumber \\
\psi _2 (\textstyle{\frac12}c )&=\e^{i (\theta_{2}-\theta_{1}) } \chi _{4} (-\textstyle{\frac12}c ),\; &
 \psi _2' (\textstyle{\frac12}c )&=\e^{i (\theta_{2}-\theta_{1}) } \chi _{4}' (-\textstyle{\frac12}c ),\end{align}
 % -------------- %
referring to the parameters $\theta_1,\theta_2\in[-\pi,\pi)$; for simplicity we will speak of them as of quasimomentum components, even if the true quasimomentum is $\frac{1}{d}(\theta_1,\theta_2)$. In addition, the functions have to be matched smoothly at the segment midpoints, that is,
 % -------------- %
\begin{align}\label{KagMidPoints}
\chi _2(0)&=\psi _4(0),\; & \chi _2'(0)&=\psi _4'(0),
\nonumber \\
\varphi _3(0)&=\psi _3(0),\; & \varphi _3'(0)&=\psi _3'(0),
\nonumber \\
\varphi _2(0)&=\chi _3(0),\; & \varphi _2'(0)&=\chi _3'(0).
\end{align}
 % -------------- %
Most important, we have to impose the matching conditions \eqref{coupling} at the vertices of graph cell. Remembering that the derivatives have to be taken in the outward direction, we get
 % -------------- %
\begin{align}\label{KagEqs}
&\psi _2(0)-\psi _1(0)  +i \ell  \left(\psi _2'(0)+\psi _1'(0)\right)=0,
\nonumber \\
&\psi _3(\textstyle{\frac12}b)-\psi _2(0)  +i \ell  \left(-\psi _3'(\textstyle{\frac12}b)+\psi _2'(0)\right)=0,
\nonumber \\
&\psi _4(\textstyle{\frac12}b)-\psi _3(\textstyle{\frac12}b) -i \ell  \left(\psi _4'(\textstyle{\frac12}b)+\psi _3'(\textstyle{\frac12}b)\right)=0,
\nonumber \\
&\psi _1(0)-\psi _4(\textstyle{\frac12}b)  +i \ell  \left(\psi _1'(0)-\psi _4'(\textstyle{\frac12}b)\right)=0,
\nonumber  \\[7pt]
&\varphi _2(-\textstyle{\frac12}b)-\varphi _1(0)  +i \ell  \left(\varphi _2'(-\textstyle{\frac12}b)-\varphi _1'(0)\right)=0,
\nonumber \\
&\varphi _3(-\textstyle{\frac12}b)-\varphi _2(-\textstyle{\frac12}b)  +i \ell  \left(\varphi _3'(-\textstyle{\frac12}b)+\varphi _2'(-\textstyle{\frac12}b)\right)=0,
\nonumber \\
&\varphi _4(0)-\varphi _3(-\textstyle{\frac12}b) +i \ell  \left(-\varphi _4'(0)+\varphi _3'(-\textstyle{\frac12}b)\right)=0,
 \\
&\varphi _1(0)-\varphi _4(0)  -i \ell  \left(\varphi _1'(0)+\varphi _4'(0)\right)=0,
\nonumber  \\[7pt]
&\chi _2(-\textstyle{\frac12}b)-\chi _1(0)  +i \ell  \left(\chi _2'(-\textstyle{\frac12}b)+\chi _1'(0)\right)=0,
\nonumber \\
&\chi _3(\textstyle{\frac12}b)-\chi _2(-\textstyle{\frac12}b)  +i \ell  \left(-\chi _3'(\textstyle{\frac12}b)+\chi _2'(-\textstyle{\frac12}b)\right)=0,
\nonumber \\
&\chi _4(0)-\chi _3(\textstyle{\frac12}b) -i \ell  \left(\chi _4'(0)+\chi _3'(\textstyle{\frac12}b)\right)=0,
\nonumber \\
&\chi _1(0)-\chi _4(0)  +i \ell  \left(\chi _1'(0)-\chi _4'(0)\right)=0.
\nonumber
\end{align}
 % -------------- %
Substituting now from \eqref{KagAnsatz} into \eqref{KagEqs}, and using \eqref{KagFloq} and \eqref{KagMidPoints}, we get a system of twelve linear equations for the coefficients $B_{3}^{\pm}, B_{4}^{\pm}, C_{1}^{\pm},C_{4}^{\pm}, D_{3}^{\pm}, D_{4}^{\pm}$; computing the corresponding determinant, taking into account that $b=d-c$, and neglecting the inessential multiplicative factor $65536\,i$, we arrive at the spectral condition
 % -------------- %
\begin{equation}\label{Kag,Pos,SC,dc}
\e^{2 i \theta _2} k^9 \ell ^3\;\sin \dfrac{kc}2\sin  \dfrac{kd}2 \sin \dfrac{k(d-c)}2\;\big( \lambda _1(k) -\lambda _2 (k)\,f_{\theta }-\lambda _3 (k)\,g_{\theta } \big) =0,
\end{equation}
 % -------------- %
where
 % -------------- %
 \begin{align*}%\label{Lambdas,dc}
& \lambda_1(k):=2(k^2 \ell ^2+1)\big(4 (k^2 \ell ^2+1)^2 \big( \cos k (c+d)+\cos k (c-2 d)+2 \cos kd+\cos 2 kd  \big)
\nonumber \\
& \qquad\qquad +\left(k^4 \ell ^4+14 k^2 \ell ^2+1\right) (2 \cos kd+1)  \cos k (2 c-d)+\left(3 k^4 \ell ^4+18 k^2 \ell ^2+3\right) \nonumber \\
& \qquad\qquad +\left(5 k^4 \ell ^4+22 k^2 \ell ^2+5\right) \big(\cos k(d-c) +\cos kc \big)\big),  \\[8pt]
& \lambda_2(k):= 8 \left(k^2 \ell ^2+1\right) \left(k^2 \ell ^2-1\right)^2 \cos  \dfrac{k(d-c)}2  \cos \dfrac{kc}2  \Big(\cos  \dfrac{k(2c-d)}2 +2 \cos \dfrac{kd}2 \Big) ,
\nonumber  \\[8pt]
& \lambda_3(k):= 16 k \ell  \left(k^2 \ell ^2-1\right)^2 \sin \dfrac{k(d-c)}2 \sin \dfrac{kc}2 \sin \dfrac{k(d-2c)}2  ,\nonumber
 \end{align*}
 % -------------- %
and the quasimomentum-dependent quantities in \eqref{Kag,Pos,SC,dc},
 % -------------- %
\begin{align}\label{fgtheta}
& f_{\theta }:=\cos \theta _1+\cos (\theta _1-\theta _2)+\cos \theta _2 ,\nonumber \\
& g_{\theta }:=\sin\theta _2+\sin (\theta _1-\theta _2)-\sin \theta _1 ,\nonumber
\end{align}
 % -------------- %
range through $[-\textstyle{\frac{3}2},3]$ and $[-\textstyle{\frac{3\sqrt{3}}2},\textstyle{\frac{3\sqrt{3}}2}]$, respectively. In particular, for $d=2c$, in which case the graph is equilateral exhibiting a repeated David-star pattern, the coefficient $\lambda_3(k)$ vanishes and the spectral condition \eqref{Kag,Pos,SC,dc} reduces to
% -------------- %
 \small
\begin{eqnarray} \label{Kag,Pos,SC,d=2c}
\lefteqn{ 4 \big(k^2 \ell ^2+1\big)  \big(2 \cos kc +1\big)\sin kc\;\sin ^2 \frac{kc}{2}} \\
 & \times \left(     \left(k^4 \ell ^4+14 k^2 \ell ^2+1\right) \cos kc+\left(k^2 \ell ^2+1\right)^2 \big(2 \cos 2kc +2 \cos 3kc+1\big)-\big(\cos kc +1\big)\left(k^2 \ell ^2-1\right)^2 f_{\theta }        \right)=0 \nonumber.
\end{eqnarray}
\normalsize
 % -------------- %

\subsection{Positive spectrum}
\label{Kag,Pos,section}
 % -------------- %
According to the spectral conditions \eqref{Kag,Pos,SC,dc} and \eqref{Kag,Pos,SC,d=2c}, the positive spectrum consists of two parts:
 % -------------- %
\paragraph*{\emph{i}. Infinitely degenerate eigenvalues}
 % -------------- %
\begin{itemize}
\item In the general case, the number $k^{2}$ belongs to the spectrum for $k=\frac{2n\pi}{l}$ with $l=\{d-c,c,d\}$ and $n\in\mathbb{N}$, may or may not be embedded in the continuous spectrum. In particular, in the equilateral case, they merge into $k^2=(\tfrac{n\pi}{c})^2$ which may not be embedded in the continuous spectrum; inspecting the large bracket in \eqref{Kag,Pos,SC,d=2c} for $k=\tfrac{n\pi}{c}$, we get $-12 \pi ^2 c^{-2} n^2 \ell ^2<0$ and $3 \left(c^2+\pi ^2 n^2 \ell ^2\right)^2- f_{\theta } \left(c^2-\pi ^2 n^2 \ell ^2\right)^2+6 \pi ^2 c^{2} n^2 \ell ^2 >0$ for odd and even $n$, respectively, the latter can be easily checked in view of the inequality $\left(c^2+\pi ^2 n^2 \ell ^2\right)^2>\left(c^2-\pi ^2 n^2 \ell ^2\right)^2$.
% -------------- %
\item  In the equilateral case, the number $k^2$ belongs to the spectrum for $k=\left((-1)^{n+1}+(6 n-3) \right)\frac{\pi}{6c}$ with $n\in\mathbb{N}$, may or may not be embedded in the continuous spectrum.
% -------------- %
\item It is possible that some positive bands degenerate to the points; in the general case, this happens at $k=\ell^{-1}$ for $\{d-c,c,d\}=\ell\big((-1)^{n+1}+(6 n-3) \big)\frac{\pi}{6}$ with $n\in\mathbb{N}$; accordingly, in the equilateral case, this happens at $k=\ell^{-1}$ for $c=\ell\big((-1)^{n+1}+(6 n-3) \big)\frac{\pi}{12}$.
% -------------- %
\end{itemize}
% -------------- %

\paragraph*{\emph{ii}. Continuous bands} \mbox{}\\
\label{kag,gen,PosCont,section}
% -------------- %
Away from the flat bands mentioned above, the rest of the spectrum is continuous having a band-and-gap structure determined by vanishing of the bracket in \eqref{Kag,Pos,SC,dc}, that is
% -------------- %
\begin{equation}\label{Kag,gen,cont,sc}
\lambda _1 (k)=\lambda _2 (k) \,f_{\theta }+\lambda _3(k) \,g_{\theta }.
\end{equation}
% -------------- %
In order to describe the bands and gaps in a more explicit way, one can inspect the right-hand side of this equation as a function of two variables $(\theta_1,\theta_2)$; using the Hessian method of determining extrema of multivariate functions as well as checking the boundaries of our rectangular domain, we find that the global extrema of the function may happen at one of the points $(0,0)$ and  $\left(\pm\tfrac{2\pi}{3},\mp\tfrac{2\pi}{3}\right)$. Hence, the positive spectrum is determined by the intersection of the function $k\mapsto \lambda _1 (k) $ with the region bordered from below and above by the curves $k\mapsto \lambda^{0}(k)$ and $k\mapsto \lambda^{\pm}(k)$, where
% -------------- %
\begin{align}
& \lambda^{0}(k):=3\,\lambda _2 (k),\nonumber \\
& \lambda^{\pm}(k):= -\frac{3}{2}\left( \lambda _2 (k)\pm\sqrt{3}\,\lambda _3 (k)\right) .\nonumber
\end{align}
% -------------- %
In other words, depending on the signs of the functions $\lambda _2 (k)$ and $\lambda _3 (k)$, a number $k^2$ belongs to a spectral band if and only if
 % -------------- %
\begin{equation}\label{Kag,gen,band,con}
k\in \left\{ k: \; \lambda^{\pm}(k)\leq \lambda _1 (k)\leq \lambda^{0}(k)\quad \cup \quad\lambda^{0}(k)\leq \lambda _1 (k)\leq \lambda^{\pm}(k)\right\}.
\end{equation}
 % -------------- %
The band-and-gap pattern of the general model in dependence on $d$ and $c$ is illustrated in Figs.~\ref{fig1} and \ref{fig2}, respectively. Moreover, two other examples for specific values of $c$ and $d$ are shown in Fig.~\ref{KagGenEx}. The spectrum of the equilateral model in dependence on, and for specific values of $c$ is illustrated in Figs.~\ref{fig3} and \ref{KagEquilEx}, respectively. We see that:
% -------------- %
\begin{outline}
\1 For $ d\geq 2 \sqrt{3}\,\ell$, the positive spectrum starts at zero, and in contrast, for $ d< 2 \sqrt{3}\,\ell$, the first positive band remains separated from zero. To see that, let us inspect the behavior of the band condition \eqref{Kag,gen,band,con} for $k\rightarrow 0+$; considering the first inequality condition, and using the Taylor expansion to the second order, we arrive at
 % -------------- %
\begin{equation}\label{Kag,gen,Tayl,small,k}
0 \; \leq \;  108- 9 k^2 \left(5 c^2-5 c d+7 d^2-36 \ell ^2\right) \; \leq \; 108-9 k^2 \left(5 c^2-5 c d+3 d^2+12 \ell ^2\right),
\nonumber
\end{equation}
 % -------------- %
where we have added $36-3 k^2 \left(5 c^2-5 c d+3 d^2+12 \ell ^2\right)$ to the inequality, all the terms with a relative error $\mathcal{O}(k^4)$. Simplifying the last two parts of the inequality, we get $d^2\geq 12 \ell ^2$, hence, small values of $k$ with $ d< 2 \sqrt{3}\,\ell$ cannot belong to the spectral bands. Needless to say, it is obvious that the second inequality in \eqref{Kag,gen,band,con} does not hold for small values of $k$. Consequently, the first positive band of the equilateral model starts at zero if $ c\geq \sqrt{3}\,\ell$, otherwise, it remains separated from zero.
% -------------- %
\1 In the high-energy regime, we have two types of asymptotic behavior. To take a closer look at their structure, we rewrite the spectral condition \eqref{Kag,gen,cont,sc} in the form
 % -------------- %
\begin{equation}\label{kag,sc,higher,powers}
\alpha(k)\cdot\, k^6+\mathcal{O}(k^5)= 0, \end{equation}
 % -------------- %
where
 % -------------- %
 \small
\begin{align}\label{Kag,asy,A}
&  \alpha(k)=4 \left(\cos  \frac{k (2 c-d)}2+2 \cos   \frac{kd}2 \right)\;\times  \\
&  \qquad  \left( \left(2 \cos k (c-d)+4 \cos kd-1\right)\cos \frac{kd}2 +\cos \frac{k (2 c+d)}2  -2\,f_{\theta } \cos  \frac{kc}{2}  \cos \frac{k (c-d)}{2}  \right) \nonumber.
\end{align}
\normalsize
 % -------------- %
Hence, as $k\rightarrow\infty$, the function $\alpha(k)$ should be close to zero which results in two types of spectral bands. We begin the discussion with the general model and discuss the equilateral model as a special case:
 % -------------- %
\2 \textbf{Pairs of narrow bands in the vicinity of the roots of $\cos \textstyle{\frac{k (2 c-d)}2}+2 \cos  \textstyle{\frac{kd}2}$} \\
The bands appear in pairs centered around the points $k$ which solve the equation $\cos \textstyle{\frac{k (2 c-d)}2}+2 \cos  \textstyle{\frac{kd}2}=0$. Note that these bands do not appear in the high energy regime of the equilateral model. Indeed, the left-hand side of the mentioned equation with $d=2c$ appears as a multiplicative factor in \eqref{kag,sc,higher,powers} corresponding to the flat bands of the second bullet point in Sec.~\ref{Kag,Pos,section}.
% -------------- %
\2 \textbf{Wide bands} \\
The bands and the gaps between them grow asymptotically but not at the same rate; they correspond to those values of $k$ for which the function in the second bracket in \eqref{Kag,asy,A} is close to zero; without loss of generality, dividing the corresponding equation by $ \cos  \frac{kc}{2}  \cos \frac{k (c-d)}{2}  $ and after simple manipulations, the sufficient condition of belonging to the spectral bands for large $k$ is obtained as
 % -------------- %
\begin{equation}\label{kag,gen,high1}
0\; \leq \;   \frac54 + \frac{\cos kd \;\cos \frac{k (2c-d)}{2}+\cos\frac{3 kd}{2}}{ \cos \frac{k (2c-d)}{2}+\cos\frac{kd}{2} }  \;     \leq \;\frac94 .
\end{equation}
 % -------------- %
As mentioned in the introduction, we are interested in the probability that a randomly chosen energy lies in the spectrum, introduced in \cite{BB13} as
 % ------------- %
\begin{equation}\label{probsigma}
P_{\sigma}(H):=\lim_{K\to\infty} \frac{1}{K}\left|\sigma(H)\cap[0,K]\right|.
\end{equation}
 % ------------- %
In the general case, we are unable to find it in a closed form and as Fig.~\ref{probfig} shows it may take different values, however, for $c$ and $d$ incommensurate, the value is the same being $\approx 0.639$. There are two ways to see that. First of all, it is clear from Fig.~\ref{probfig} that for rational $\frac{c}{d} = \frac{p}{q}$ with large coprime $p$ and $q$, which we may regard as rational approximation to a given irrational number, the probability is near to the indicated value. Secondly, keeping the leading order in the band condition \eqref{Kag,gen,band,con} we get asymptotically
 % ------------- %
\begin{equation}\label{asycond}
\Big( 2\cos\frac{k(c-2d)}{2} + \cos\frac{kc}{2} \Big) \Big( \cos\frac{k(c-d)}{2} + 2\cos\frac{k(c+d)}{2} \Big) \Big( \cos\frac{k(2c-d)}{2} + 2\cos\frac{kd}{2} \Big) \ge 0
\end{equation}
 % ------------- %
We can rewrite the left-hand side of \eqref{asycond} as a function of $\sin\frac{kb}{2}$ and $\sin\frac{kc}{2}$ only. The resulting expression is quite complicated and gives no hope to solve the inequality, however, if $b$ and $c$ are incommensurate, one can regard the two sines as a pair of independent identically distributed random variables and compute numerically the probability that such a quantity will be non-negative; this yields again the value mentioned above. Indeed, if we calculate the area of the gray parts in Fig.~\ref{probfig2}, i.e. the region where the left-hand side of \eqref{asycond} is non-negative, and divide it by $4\pi^2$, we get $\approx 0.639081$.
Hence, despite the vertex coupling is in the present case substantially different, we find again the universality demonstrated in \cite{BB13} for periodic graphs with Kirchhoff vertices.

\medskip

Let us turn to the \emph{equilateral case}, $b=c$. Requiring the second bracket in \eqref{Kag,asy,A} to vanish, we get
 % -------------- %
\begin{equation}\label{kag,d=2c,high1}
 \cos ^2 \frac{kc}{2} \big(4 \cos kc-4 \cos 2kc+f_{\theta }-3 \big)=\mathcal{O}(k^{-1})     ,
\end{equation}
 % -------------- %
indicating that, again, we have two types of spectral bands:
 % -------------- %
\2[--] \textbf{Pairs of narrow bands in the vicinity of $k=(2n-1)\frac{\pi}{c}$, $n\in\mathbb{N}$} \\
The two bands around the points $k=(2n-1)\frac{\pi}{c}$ and the gap between them have asymptotically constant width as $n\rightarrow \infty$. To see that, we rewrite the expression in the large brackets in \eqref{Kag,Pos,SC,d=2c} in the asymptotic form
 % -------------- %
\begin{equation}\label{kag,d=2c,Alter,Asy}
\beta_1(k)+\frac{\beta_2(k)}{k^2}=\mathcal{O}(k^{-4}),\end{equation}
  % -------------- %
with
 % -------------- %
\begin{align}
&  \beta_1(k)= -2 \ell ^4 \cos ^2 \frac{kc}{2}  \left(4 \cos kc-4 \cos 2kc+f_{\theta }-3\right)        , \nonumber\\
&  \beta_2(k)= 2 \ell ^2 \left( \left(\cos kc+1\right)f_{\theta } +7 \cos kc+2 \cos 2kc+2 \cos 3kc+1\right)         .\nonumber \end{align}
 % -------------- %
Then, setting $k=(2n-1)\frac{\pi}{c}+\delta$, we get $k^{-2}=\frac{c^2}{4n^2\pi^2}+\mathcal{O}(n^{-3})$ as $n\to\infty$. Substituting these into \eqref{kag,d=2c,Alter,Asy} and solving the resulting equation for $\delta$, we obtain
 % -------------- %
\[\delta=\frac{\sqrt{6}}{\pi  \ell  \sqrt{11-f_{\theta }}}\,\frac{1}{n}+\mathcal{O}(n^{-3}) .\]
 % -------------- %
Since the band edges correspond to $f_{\theta }=-\tfrac{3}{2}$ and $3$, the width of the bands and the gap between them are (on the energy scale) respectively determined as $\frac{2}{5c\ell} \sqrt{3}+\mathcal{O}(n^{-1})$ and $\frac{16}{5c\ell} \sqrt{3}+\mathcal{O}(n^{-1}) $ as $n\rightarrow \infty$.
 % -------------- %
\2[--] \textbf{Wide bands} \\
Both bands and gaps grow asymptotically, again, not at the same rate; manipulating the expression in the bracket in \eqref{kag,d=2c,high1}, or equivalently, substituting $d=2c$ in \eqref{kag,gen,high1}, we find that large values of $k$ belong to the spectral bands if and only if
 % -------------- %
\begin{equation}\label{kag,bandCondition,high,d=2c}
0\leq \xi(k) \leq\frac{9}{8}  \; ; \quad  \xi(k):=\cos kc-\cos 2kc,
\end{equation}
 % -------------- %
with a relative error $\mathcal{O}(k^{-1})$. The function $\xi(k)$ is periodic with the period $T=\tfrac{2\pi}{c}$ and one can easily check that the maximum value of this function is $\tfrac{9}{8}$ which happens at $k= \tfrac {1}{c} | 2m\pi   \pm \arcsec 4 |$ with $m\in\mathbb{Z}$.

It remains to calculate the probability that $\xi(k)$ is positive for a randomly chosen value of $k$. The roots of $\xi(k)$ in the period are $\tfrac{2\pi}{3c}$ and $\tfrac{4\pi}{3c}$. On the other hand, we have $\xi(\tfrac{\pi}{c})=-2$, hence, $\xi(k)$ is negative over the domain $\left(\tfrac{2\pi}{3c},\tfrac{4\pi}{3c}\right)$ and thus, the probability \eqref{probsigma} is for any $c$ equal to
 % -------------- %
\begin{equation}\label{probequi}
P_{\sigma}(H)=1-\frac{1}{T}\left( \frac{4\pi}{3c}-\frac{2\pi}{3c} \right)=\frac{2}{3}.
\end{equation}
% -------------- %
This differs from the universal value obtained above which is not surprising: the equilateral character of the graph, as a particular case of commensurability, means that the ergodicity of the flow which was crucial for the universality \cite{BB13} is lost.

 \1 The fact that the lattice exhibits a nonvanishing transport in the sense of the probability \eqref{probsigma} follows from the fact that the high-energy limit of the vertex scattering matrix is nontrivial. In fact, for vertices of degree four we get from \eqref{sij,onshell} that
% -------------- %
\[ \underset{k\to \infty }{\text{lim}}S(k)= {\scriptsize
\frac12 \left(\begin{array}{cccc}
 \phantom{-}1 &     \phantom{-}1 &       -1 &                  \phantom{-}1 \\
 \phantom{-}1 &     \phantom{-}1 &        \phantom{-}1 &     -1 \\
 -1 &               \phantom{-}1 &        \phantom{-}1 &       \phantom{-}1 \\
 \phantom{-}1 &     -1 &                  \phantom{-}1 &       \phantom{-}1 \\
\end{array}\right)  },\]
% -------------- %
which means that the probabilities of leaving the vertex in any of the four direction are asymptotically the same. A comparison with other lattices and chains with vertices of degree four \cite{BET21, ET18} shows, however, that the quantity \eqref{probsigma} is different in different situations depending on the topology of the structure.
% -------------- %
\1 While generically the gaps are open, it may happen that some of them close for some particular values of the parameters. This happens when the boundaries of neighboring bands touch as illustrated in Fig.~\ref{fig1}; note that such crossing points may occur in sequences with the same energy. In general, it is not easy to find their coordinates in a closed form, however, introducing the symbol $\Delta:=\big( \lambda _1(k) -\lambda _2 (k)\,f_{\theta }-\lambda _3 (k)\,g_{\theta } \big)$ for the bracket in \eqref{Kag,Pos,SC,dc} we can identify such situations using the sufficient condition,
 % ------------- %
\begin{equation}\label{cross,con}
\frac{\partial \Delta}{\partial \theta_1}=\frac{\partial \Delta}{\partial \theta_2}=\frac{\partial \Delta}{\partial k}=\frac{\partial \Delta}{\partial d}=0,
\end{equation}
 % ------------- %
supposing that $c$ and $\ell$ are fixed. For the first two derivatives we get the expressions
 % ------------- %
\begin{align}
&   \frac{\partial \Delta}{\partial \theta_1}=\sin \Big(\theta _1-\frac{\theta _2}{2}\Big) \left(\lambda _2(k) \cos \frac{\theta _2}{2}-\lambda _3(k) \sin \frac{\theta _2}{2}\right), \nonumber\\
&   \frac{\partial \Delta}{\partial \theta_2}=\sin \Big(\theta _2-\frac{\theta _1}{2}\Big) \left(\lambda _2(k) \cos \frac{\theta _1}{2}+\lambda _3(k) \sin \frac{\theta _1}{2}\right),\nonumber
\end{align}
 % ------------- %
which vanish at $(\theta_1,\theta_2)=(0,0)$ and $(\pm\tfrac{2\pi}{3},\mp\tfrac{2\pi}{3})$ corresponding to the band edges. Inspecting then the last two derivatives in \eqref{cross,con} at these values, one obtains a system of two equations in variables $k$ and $d$ that may always be fulfilled for particular values of parameters.
% -------------- %
\1 The spectral bands are symmetric with respect to the exchange of $c$ to $d-c$, as seen in Fig.~\ref{fig2}. The band condition \eqref{Kag,gen,band,con} depends on the three functions $\lambda_1(k)$, $\lambda_2(k)$ and $\lambda_3(k)$; the invariance of the first two functions under $c \leftrightarrow  d-c$ are easily checked since $\cos$ is an even function; in the case of $\lambda_3(k)$, although it is an odd function, the functions $\lambda^{\pm}(k)$ in the band condition \eqref{Kag,gen,band,con} contain it with both the positive and negative signs.
 % -------------- %
\1 If the size of the lattice cell is large, that is, in the asymptotic regime $d\rightarrow\infty$, the the number of bands in a fixed energy interval increases, roughly linearly with $d$, however, the probability to belong to the spectrum remains asymptotically the same.
 % -------------- %
\end{outline}
 % -------------- %

\subsection{Negative Spectrum}
\label{KagNeg,section}
 % -------------- %
Replacing the momentum variable $k$ in \eqref{Kag,Pos,SC,dc} by $i\kappa$ with $\kappa>0$, we arrive at the spectral condition
 % -------------- %
\begin{equation}\label{Kag,Neg,SC,dc}
\e^{2 i \theta _2} \kappa^9 \ell ^3\;\sinh \frac{\kappa c}2\sinh \frac{\kappa d}2 \sinh \frac{\kappa (d-c)}2\;\Big(  \tilde{\lambda} _1 (\kappa ) -\tilde{\lambda} _2 (\kappa )\,f_{\theta }-\tilde{\lambda} _3(\kappa )\,g_{\theta } \Big) =0, \end{equation}
 % -------------- %
where
 % -------------- %
 \begin{align*} %\label{Neg,Lambdas,dc}
& \tilde{\lambda} _1 (\kappa ) :=2(1-\kappa^2 \ell ^2)\bigg(4 (\kappa^2 \ell ^2-1)^2 \Big( \cosh \kappa (c+d)+\cosh \kappa (c-2 d)+2 \cosh \kappa d+\cosh 2 \kappa d  \Big)
\nonumber \\
& \qquad\qquad +\left(\kappa ^4 \ell ^4-14 \kappa ^2 \ell ^2+1\right) (2 \cosh \kappa d+1)  \cosh \kappa (2 c-d)+\left(3 \kappa^4 \ell ^4-18 \kappa^2 \ell ^2+3\right)
\nonumber \\
& \qquad\qquad +\left(5 \kappa^4 \ell ^4-22 \kappa^2 \ell ^2+5\right) \big(\cosh \kappa (d-c) +\cosh \kappa c \big)\bigg),  \\[8pt]
& \tilde{\lambda} _2 (\kappa ) := 8 \left(1-\kappa^2 \ell ^2\right) \left(\kappa^2 \ell ^2+1\right)^2 \left(\cosh   \frac{\kappa(2c-d)}2 +2 \cosh  \frac{\kappa d}2 \right)\cosh  \frac{\kappa(d-c)}2  \cosh \frac{\kappa c}2   ,
\nonumber  \\[8pt]
& \tilde{\lambda} _3 (\kappa ) := 16 \kappa \ell  \left(\kappa^2 \ell ^2+1\right)^2 \sinh \frac{\kappa(d-c)}2 \sinh \frac{\kappa c}2 \sinh \frac{\kappa(d-2c)}2 .\nonumber
 \end{align*}
 % -------------- %
Except for the equilateral case, flat bands in the negative part of the spectrum are obviously absent. Mimicking the argument of Sec.~\ref{Kag,Pos,section}, we infer that a number $-\kappa^2$ belongs to a spectral band if and only if
 % -------------- %
\begin{equation}\label{Kag,gen,Neg,band,con}
\kappa\in \left\{ \kappa: \; \tilde{\lambda}^{\pm}(\kappa)\leq \tilde{\lambda} _1 (\kappa)\leq \tilde{\lambda}^{0}(\kappa)\quad \cup \quad\tilde{\lambda}^{0}(\kappa)\leq \tilde{\lambda} _1 (\kappa)\leq \tilde{\lambda}^{\pm}(\kappa)\right\},
\end{equation}
 % -------------- %
where $\tilde{\lambda}^{0,\pm}(\kappa)$ are defined as
 % -------------- %
\begin{align}
& \tilde{\lambda}^{0}(\kappa):=3\,\tilde{\lambda} _2 (\kappa),\nonumber \\
& \tilde{\lambda}^{\pm}(\kappa):= -\frac{3}{2}\left( \tilde{\lambda} _2 (\kappa)\pm\sqrt{3}\,\tilde{\lambda} _3 (\kappa)\right) .\nonumber
\end{align}
% -------------- %
In the equilateral case, $\tilde{\lambda} _3 (\kappa )$ in \eqref{Kag,Neg,SC,dc} vanishes and the spectral condition reduces to
% -------------- %
 \small
\begin{eqnarray} \label{Kag,neg,SC,d=2c,tot}
\lefteqn{ 4  \big(\kappa ^2 \ell ^2-1 \big)  (2 \cosh \kappa c +1) \sinh \frac{\kappa c}{2} \sinh \kappa c \;\times } \\
 & \left(  \left(\kappa ^2 \ell ^2-1\right)^2 (2 \cosh2\kappa c+2 \cosh3\kappa c+1)+\left(\kappa ^4 \ell ^4-14 \kappa ^2 \ell ^2+1\right) \cosh\kappa c   - \left(\kappa ^2 \ell ^2+1\right)^2 (\cosh \kappa c+1)f_{\theta }    \right)=0 \nonumber.
\end{eqnarray}
\normalsize
 % -------------- %
The negative spectrum of the general and the equilateral lattices in dependence on $d$ and $c$ are shown in Figs.~\ref{fig1}, \ref{fig2} and \ref{fig3}, respectively. Concerning the number of negative bands, since the elementary cell of the kagome lattice contains three vertices of degree four, the corresponding matrix $U$ in each of them has one eigenvalue in the upper complex halfplane-- cf. eq.~\eqref{Neg,Eig,StarG} -- thus the negative spectrum cannot have more than three bands in accordance with Theorem 2.6 of \cite{BET21}. It should be noted that in the general case some gaps may close at specific values of $\kappa$. The crossing points are given by a relation analogous to \eqref{cross,con}; it again does not allow for solution in a closed form but one can check that such crossings indeed happen at the quasimomentum component values $\theta_1=-\theta_2=\pm\tfrac{2\pi}{3}$ corresponding to the band edges. Here, we see that:
% -------------- %
\begin{outline}
\1  As indicated, a flat band band occurs only in the equilateral case, corresponding to the energy $-\ell^{-2}$; it may not be embedded in the continuous spectrum as we will see below.
% -------------- %
\1 The number $-\ell^{-2}$ always belongs to the spectrum. Inspecting \eqref{Kag,Neg,SC,dc} for $\kappa=\ell^{-1}$, we get
\[-64  \; g_{\theta }\, \sinh \frac{c}{2 \ell } \sinh \frac{d-2 c}{2 \ell } \sinh \frac{d-c}{2 \ell } =0,\]
% -------------- %
which holds for $g_{\theta }=0$ independently of other parameters. In particular, as mentioned above, for $d=2c$, the band containing this number shrinks to the point $-\ell^{-2}$ corresponding to the flat band of the equilateral model.
% -------------- %
\1  For $ d\leq 2 \sqrt{3}\,\ell$, the first negative band reaches zero, while for $ d> 2 \sqrt{3}\,\ell$ the negative spectrum remains separated from zero. Following an argument similar to that of Sec.~\ref{Kag,Pos,section}, using the Taylor expansion around $\kappa\rightarrow 0+$ in \eqref{Kag,gen,Neg,band,con}, one can easily check that small values of $\kappa$ with $ d> 2 \sqrt{3}\,\ell$ do not correspond to the spectral points. In particular, in the equilateral case, the first negative band reaches zero if $ c\leq  \sqrt{3}\,\ell$, otherwise it remains separated from zero.
 % -------------- %
 \1 The negative bands are symmetric with respect to the interchange of $c$ and $d-c$, cf. Fig.~\ref{fig2}. This property can be justified by an argument similar to that used in Sec.~\ref{Kag,Pos,section} for the mirror symmetry of the positive bands.
 % -------------- %
 \1 It is also interesting to inspect the situation where one of the edges, as well as the scale parameter $\ell$, is kept fixed while the other becomes large assuming, say, that $d\gg2c$. Since attraction responsible for the negative spectrum comes from the vertex coupling only and the transport requires tunneling over the edges which are now classically forbidden zones, one expects that negative bands shrink to points as $d\to\infty$. It is indeed the case; the see that we rewrite the spectral condition \eqref{Kag,Neg,SC,dc} in the form
% -------------- %
\begin{equation}\label{kag,large,d,gen}
 \left(1-\kappa ^2 \ell ^2\right)f(\ell,c ;\kappa )\,\e^{2\kappa d }+ g(\ell,c ;\kappa ,f_{\theta },g_{\theta })\,\e^{\kappa d }+h(\ell,c ;\kappa )+\mathcal{O}(\e^{-\kappa d })=0 ,
\end{equation}
% -------------- %
with
% -------------- %
\begin{align}
& f(\ell,c ;\kappa ):=  4 \left(\e^{- \kappa c }+1\right) \left(\kappa ^2 \ell ^2-1\right)^2+\e^{-2 \kappa c } \left(\kappa ^4 \ell ^4-14 \kappa ^2 \ell ^2+1\right)     ,
\nonumber\\[8pt]
& g(\ell,c ;\kappa ,\theta_1,\theta_2):= \left(\e^{-2 \kappa c }+3 \e^{-\kappa c }+2\right)  \left(\kappa ^2 \ell ^2-1\right) \left(\kappa ^2 \ell ^2+1\right)^2\,f_{\theta }-2 \kappa  \ell  \e^{-2 \kappa c } \left(\e^{\kappa c }-1\right) \left(\kappa ^2 \ell ^2+1\right)^2 \,g_{\theta }  \nonumber\\
& \quad -\left(\kappa ^2 \ell ^2-1\right) \left(\e^{-2 \kappa c } \left(\kappa ^4 \ell ^4-14 \kappa ^2 \ell ^2+1\right)+\e^{-\kappa c } \left(5 \kappa ^4 \ell ^4-22 \kappa ^2 \ell ^2+5\right)+4 \left(\e^{\kappa c }+2\right) \left(\kappa ^2 \ell ^2-1\right)^2\right),
\nonumber \\[8pt]
& h(\ell,c ;\kappa ):= \left(1-\kappa ^2 \ell ^2\right) \left(\e^{-\kappa c} \left(\e^{2 \kappa c}+1\right) \left(5 \kappa ^4 \ell ^4-22 \kappa ^2 \ell ^2+5\right)+6 \left(\kappa ^4 \ell ^4-6 \kappa ^2 \ell ^2+1\right)\right) .
 \nonumber \end{align}
% -------------- %
For large $d$ the bands are thus in the vicinity of zeros of the first, $\theta$-independent term, being exponentially narrow with respect to $d$. One of those limit points is $-\ell^{-2}$, and one of the others in each of the the intervals $(0,\ell^{-1})$ and $(\ell^{-1},\infty)$ being determined by the condition $f(\ell,c ;\kappa )=0$. To see that, note that $f(\ell,c;\cdot)$ is continuous assuming the values $9$ and $-12 \e^{-\frac{2 c}{\ell }}$ at $\kappa=0,\, \ell^{-1}$, respectively, and $\lim_{\kappa\to\infty} f(\ell,c ;\kappa )=+\infty$; recall that by general principles the kagome lattice cannot have more than three negative bands.
% -------------- %
\1 Consider finally the equilateral case in the asymptotic regime $c\rightarrow\infty$. We know that $-\ell^{-2}$ is a flat band, the other two bands now shrink to the same value. To see that, we rewrite the spectral condition \eqref{Kag,neg,SC,d=2c,tot} in the form
% -------------- %
\begin{equation}\label{kag,Neg,d=2c,sc}
f_{\theta }=F(\kappa):= \frac{\left(\kappa ^2 \ell ^2-1\right)^2 (2 \cosh 2\kappa c +2 \cosh 3 \kappa c +1)+\left(\kappa ^4 \ell ^4-14 \kappa ^2 \ell ^2+1\right) \cosh \kappa c}{\left(\kappa ^2 \ell ^2+1\right)^2 (\cosh\kappa c+1)} .
\end{equation}
% -------------- %
We note that $F(\ell^{-1})=-\tfrac{3}{2}\big(\tanh ^2 \tfrac{c}{2 \ell } +1\big)<-\tfrac{3}{2}$ since $\tanh x>0$ holds for $x>0$, hence the flat band is not embedded in the continuous spectrum. On the other hand, we have $F(0)=3$ and $\lim_{\kappa\to\infty} F(\kappa)=+\infty$, hence there is one negative band below and above the energy $-\ell^{-2}$. For large values of $c$, the spectral condition \eqref{kag,Neg,d=2c,sc} reads $ \left(\kappa ^2 \ell ^2-1\right)^2 \e^{3\kappa c}+\mathcal{O}(\e^{2\kappa c})=0 $ implying again that the bands shrink exponentially fast. To estimate the band widths, we put $\kappa^2=\ell^{-2}+\varepsilon$ in \eqref{Kag,neg,SC,d=2c,tot} obtaining
 % -------------- %
 \[   \varepsilon=\pm  \sqrt{2 f_{\theta }+6}\,\ell^{-2}\e^{-\frac{c}{\ell }}+\mathcal{O}(\e^{-\frac{2c}{\ell }}).   \]
% -------------- %
This yields the following asymptotic expression for the width of the two bands,
 % -------------- %
\[  \Delta E= \sqrt{3}\ell^{-2} \e^{-\frac{c}{\ell }}+\mathcal{O}( \e^{-2\frac{c}{\ell }} ); \]
% -------------- %
this behavior is seen in Fig.~\ref{fig3}.
\end{outline}
 % -------------- %
%\pagebreak
 % -------------- %

\section{Triangular lattice}
\label{Trig,section}
\setcounter{equation}{0}

\begin{figure}[!htb]
\centering
\includegraphics[scale=0.5]{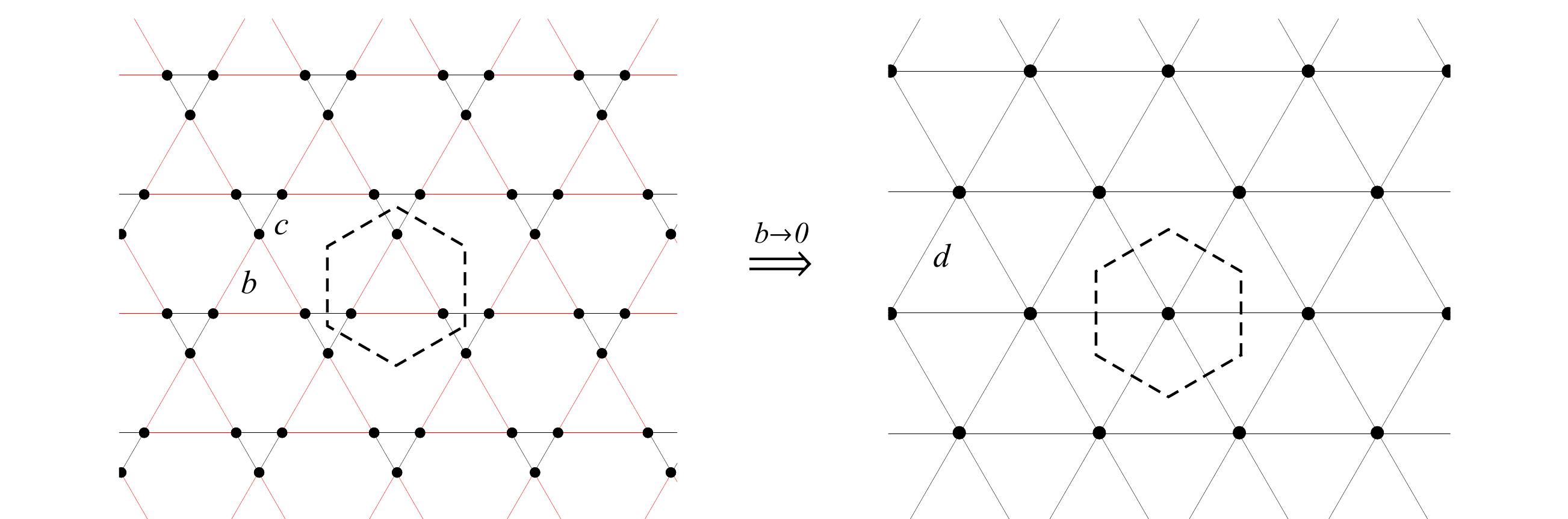}
\caption{The kagome lattice degenerates to the triangular one when one of the edge lengths shrinks to zero assuming that $d=b+c$ is fixed.}
\label{KagTendTrig}
\end{figure}

\begin{figure}[!htb]
\centering
\includegraphics[scale=1.2]{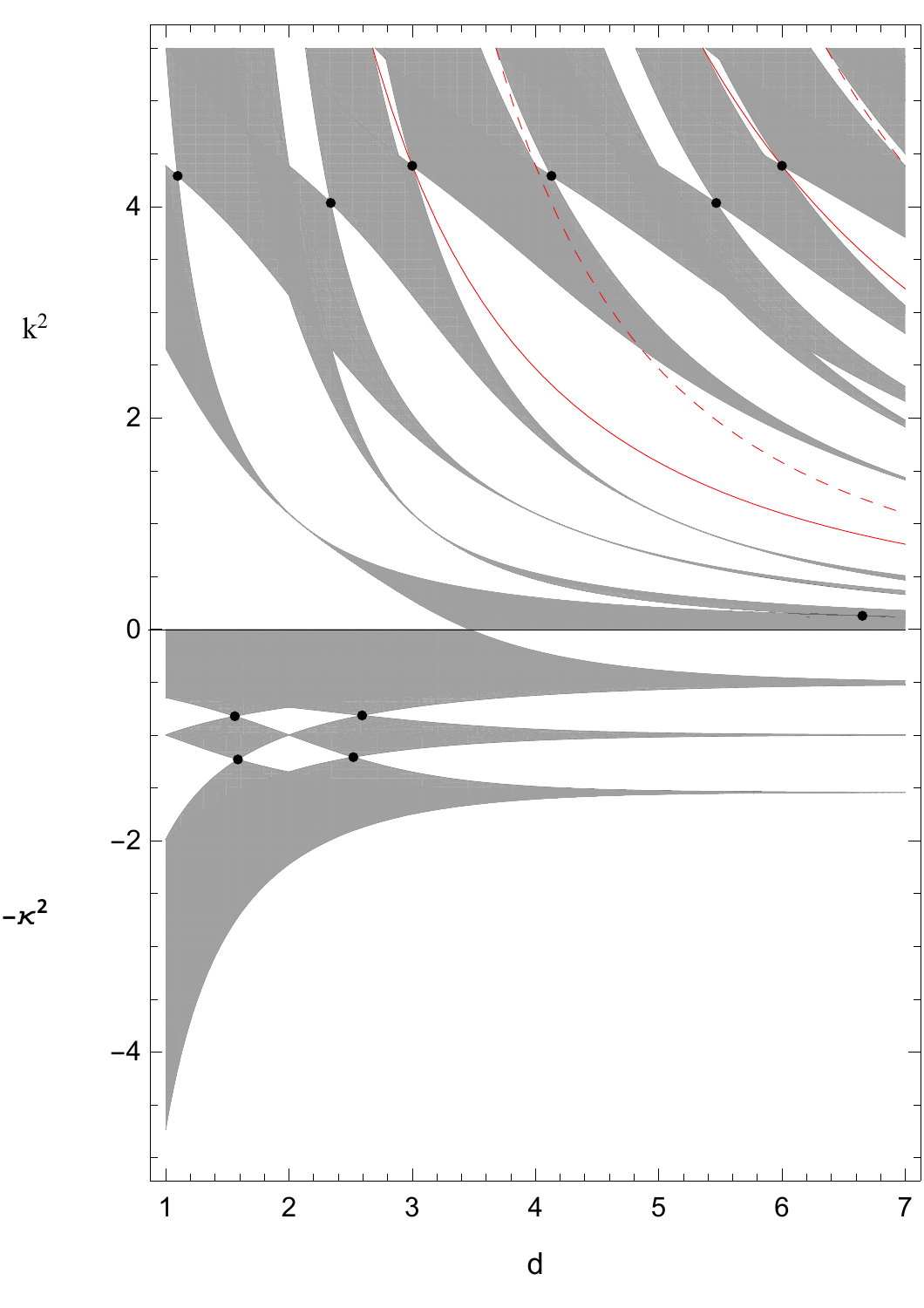}
\caption{ Spectrum of the general kagome lattice in dependence on $d$ for $c=\ell=1$. The band edge crossings is indicated by black dots. The red curves correspond to the flat bands $\frac{4 n^2\pi^2}{d^2}$ and $\frac{4 n^2\pi^2}{(d-1)^2}$ (solid and dashed, respectively) with $n=1,2$; the flat bands $4 n^2\pi^2$ lay outside the picture area. On the other hand, the values $d=\left \{   \frac{2\pi}{3},\, \frac{2\pi}{3}+1, \, \frac{4\pi}{3}, \, \frac{4\pi}{3}+1   \right\}$ with energy $k^2=1$ correspond to the degenerate eigenvalues of the third bullet point in Sec.~\ref{Kag,Pos,section}.}
\label{fig1}
\end{figure}

\begin{figure}[!htb]
\centering
\includegraphics[scale=1.1]{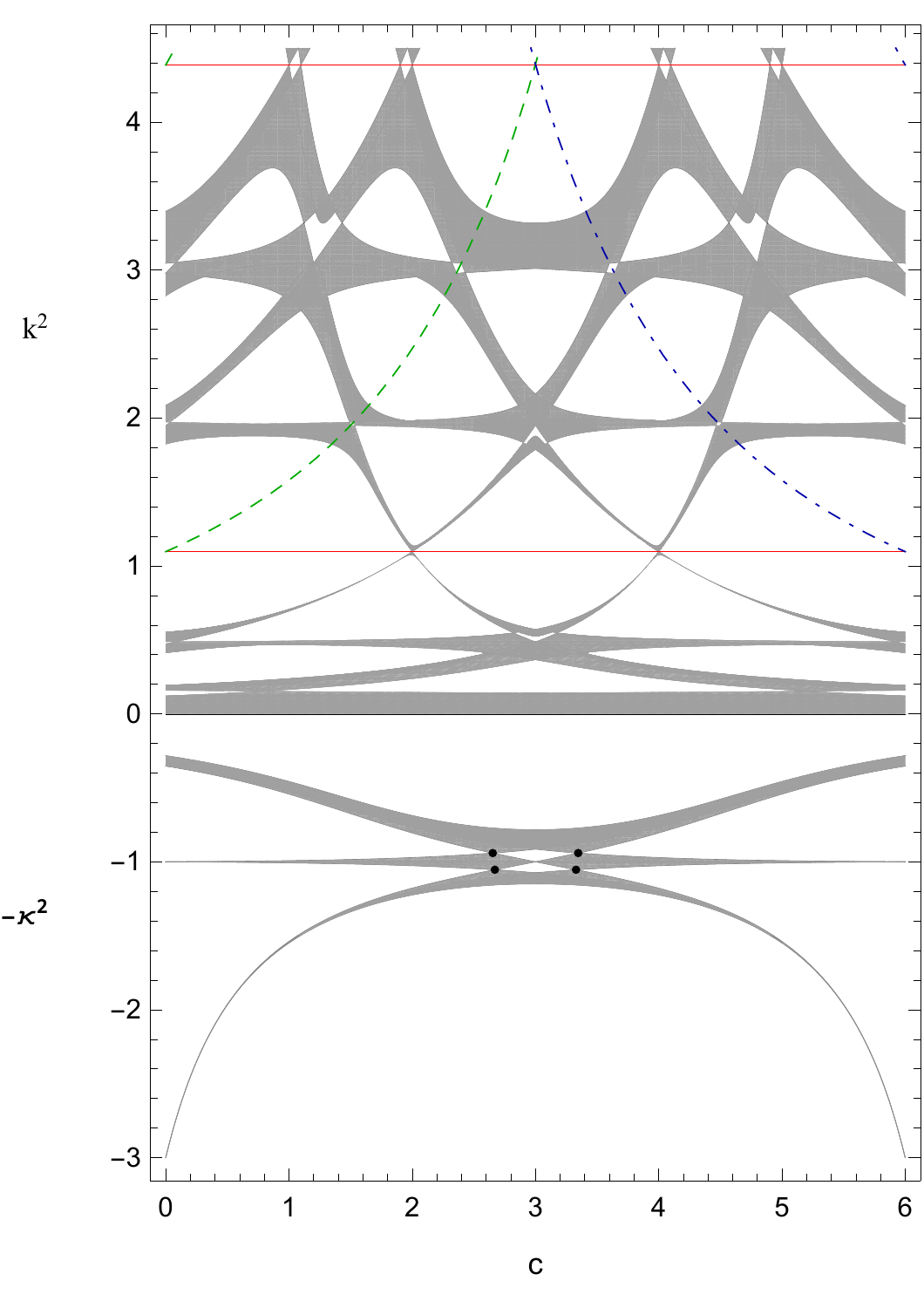}
\caption{ Spectrum of the general kagome lattice in dependence on $c$ for $d=6$ and $\ell=1$. Since $d>2\sqrt{3}$, the positive spectrum starts at zero while the negative spectrum remains separated from zero. The red solid lines correspond to the flat bands $\frac{4n^2\pi^2}{d^2}$ with $n=1,2$. The green dashed curves correspond to the flat bands $\frac{4n^2\pi^2}{(d-c)^2}$ with $n=1,2$. The blue dot-dashed curves correspond to the flat bands $\frac{4n^2\pi^2}{c^2}$ with $n=1,2$. On the other hand, the values $c=\left \{   6-\frac{4\pi}{3},\, \frac{2\pi}{3}, \, 6-\frac{2\pi}{3} , \frac{4\pi}{3} \right\}$ with energy $k^2=1$ correspond to the degenerate eigenvalues of the third bullet point in Sec.~\ref{Kag,Pos,section}; this is not well seen in the picture but magnifying it one can check that the bands indeed shrink to points there.}
\label{fig2}
\end{figure}

\begin{figure}[!htb]
  \centering
  \subfloat[Spectral bands for $d=3$; the first band remains separated from zero.]{\includegraphics[scale=1.1]{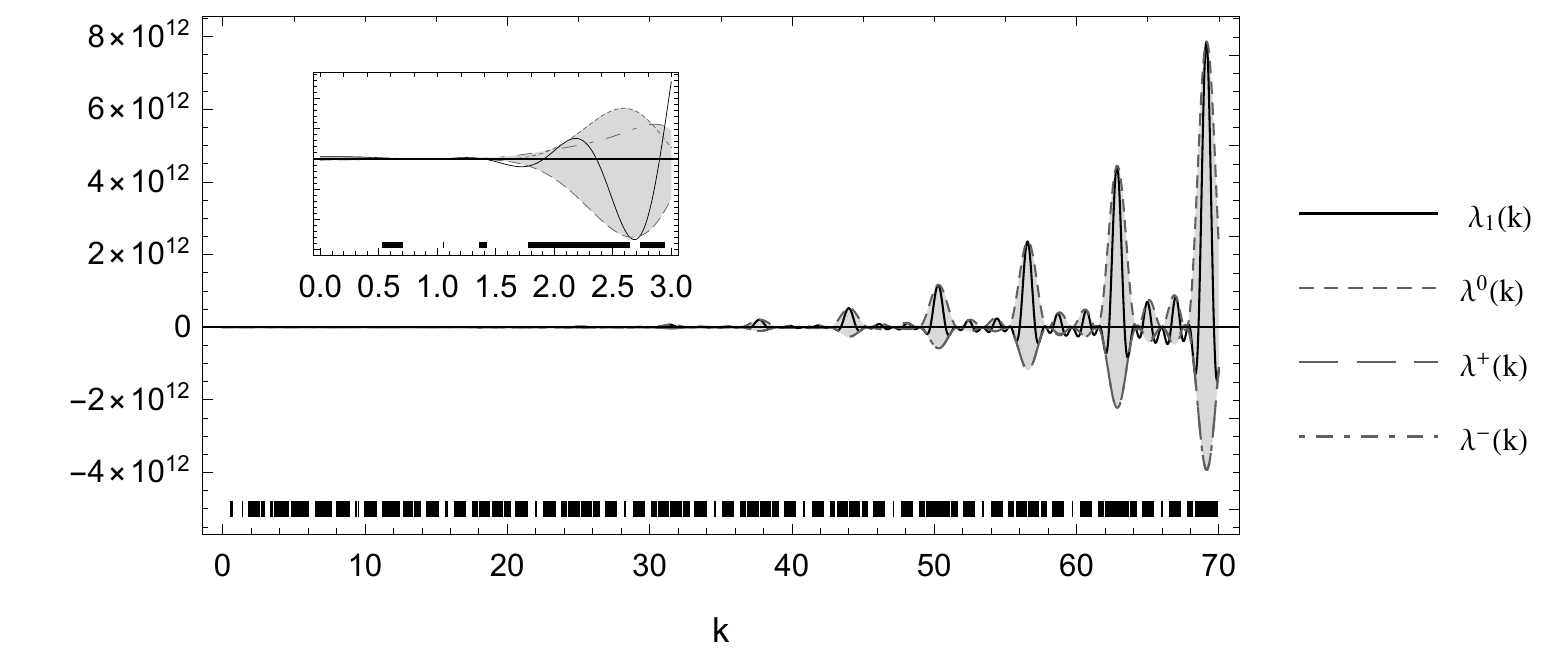}
  \label{KagGenEx1}}
  \hfill
  \subfloat[Spectral bands for $d=4$; the first band starts at zero.]{\includegraphics[scale=1.1]{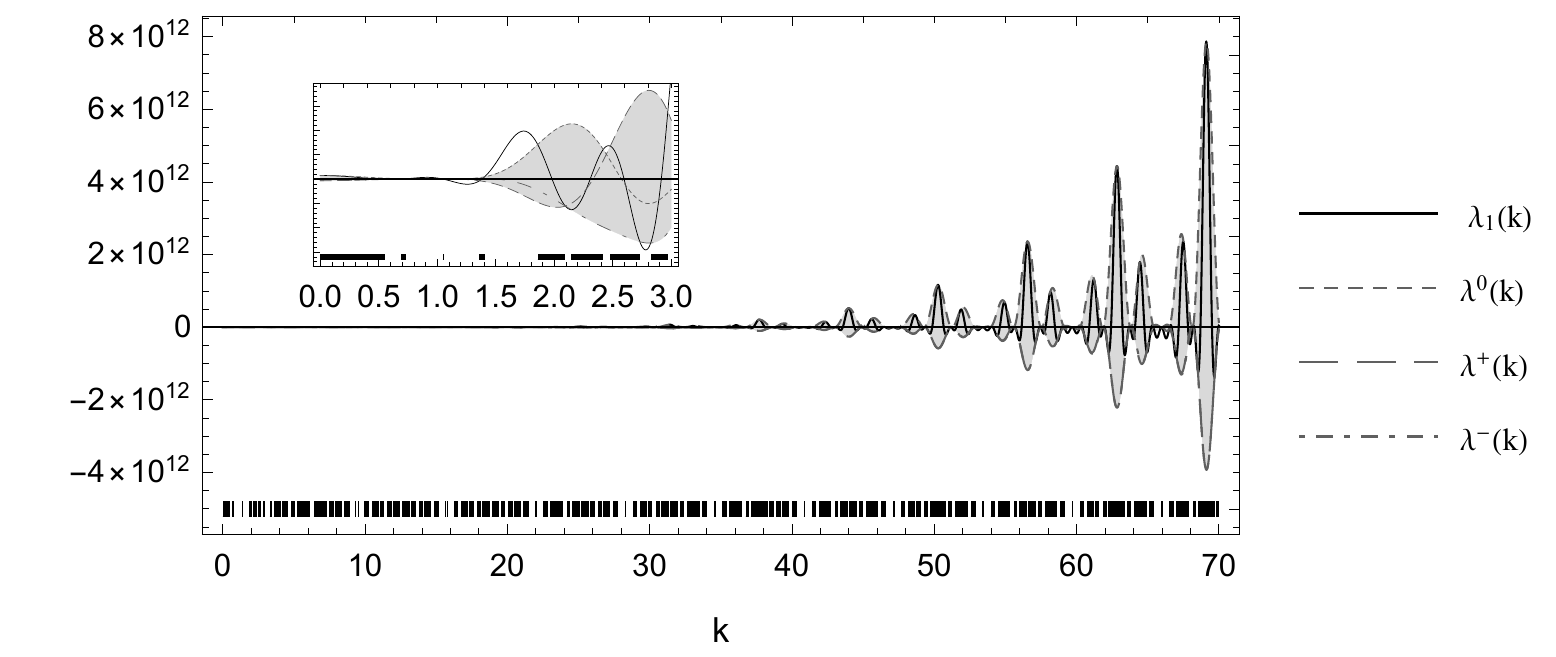}
  \label{KagGenEx2}}
\caption{Spectral condition \eqref{Kag,gen,band,con} for $d=3,4$ with $c=\ell=1$. }
\label{KagGenEx}
\end{figure}

\begin{figure}[!htb]
\centering
\includegraphics[scale=1.2]{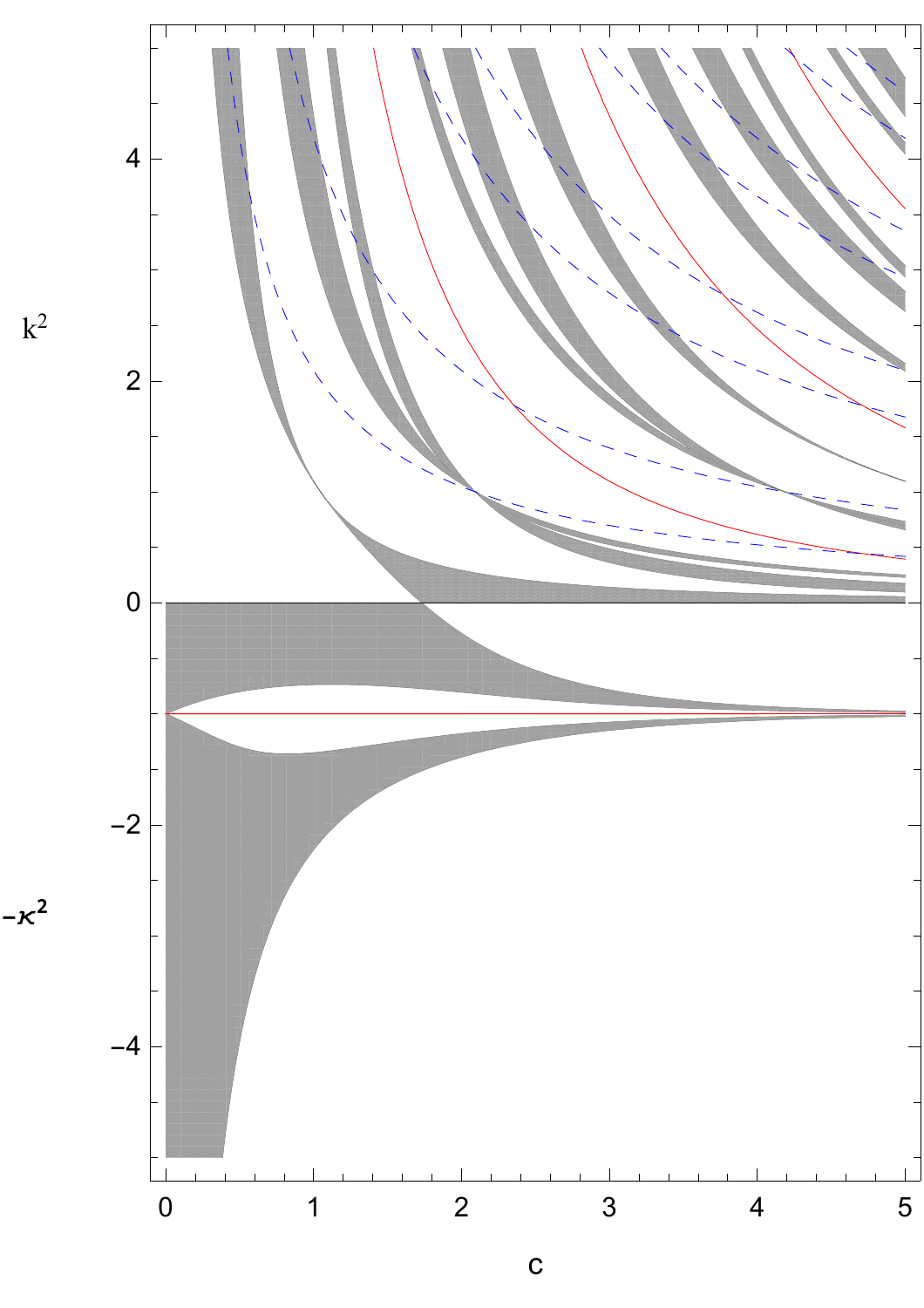}
\caption{ Spectrum of the equilateral kagome lattice in dependence on $c$ for $\ell=1$. The red solid curves in the positive spectrum correspond to the flat bands $\frac{n^2\pi^2}{c^2}$ with $n=1,2,3$ respectively. The blue dashed curves correspond to the flat bands $k^2$ for $ k=\left((6 n-3)+(-1)^{n+1}\right)\tfrac{\pi }{6 c} $ with $n=1,\ldots,8$ respectively. On the other hand, the values $c=\left \{   \frac{\pi}{3},\, \frac{2\pi}{3}, \, \frac{4\pi}{3}  \right\}$ with energy $k^2=1$ correspond to the degenerate eigenvalues of the third bullet point in Sec.~\ref{Kag,Pos,section}. The red line in the negative spectrum corresponds to the flat band $-\ell^{-2}$.}
\label{fig3}
\end{figure}

\begin{figure}[!htb]
  \centering
  \subfloat[Spectral bands for $c=1$; the first band remains separated from zero.]{\includegraphics[scale=1.1]{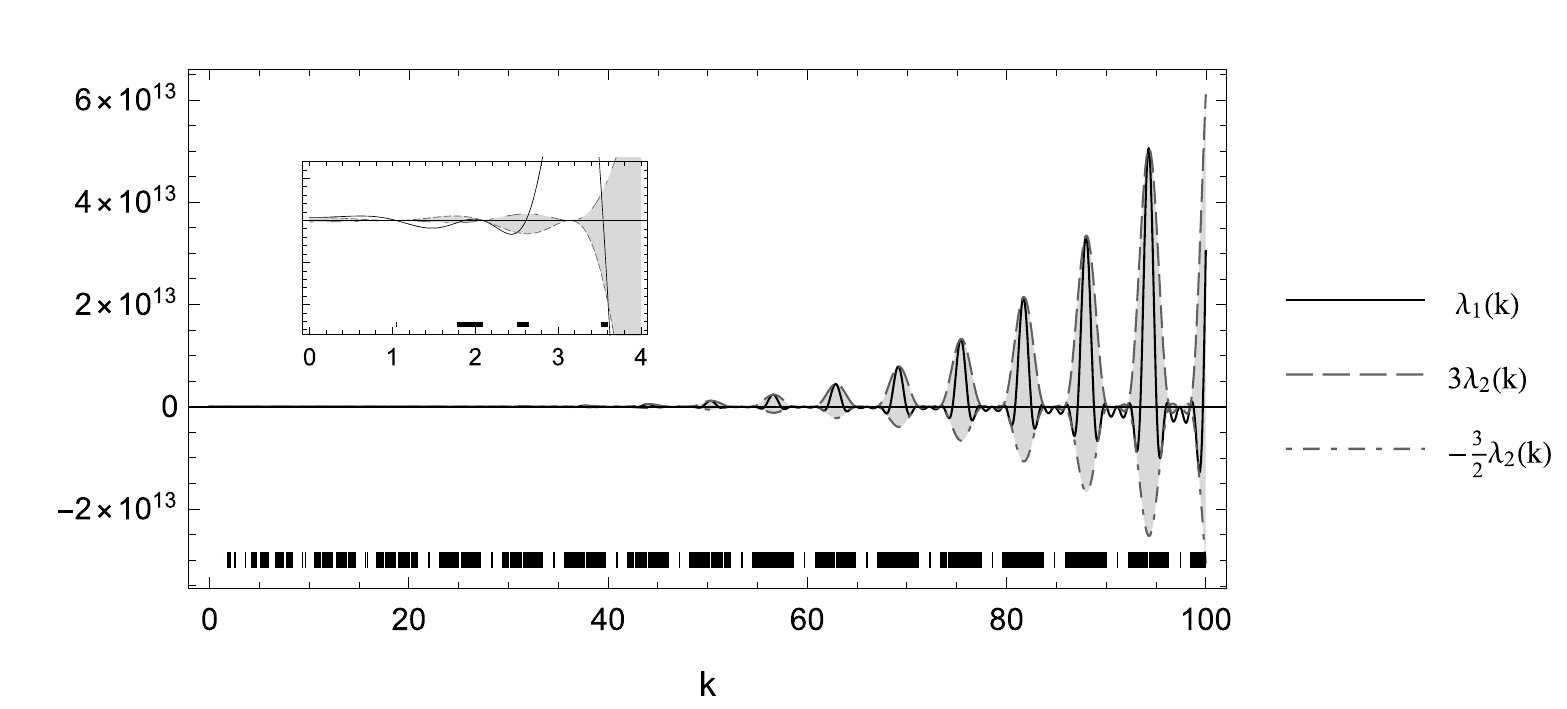}
  \label{KagEquilEx1}}
  \hfill
  \subfloat[Spectral bands for $c=3$; the first band starts at zero.]{\includegraphics[scale=1.1]{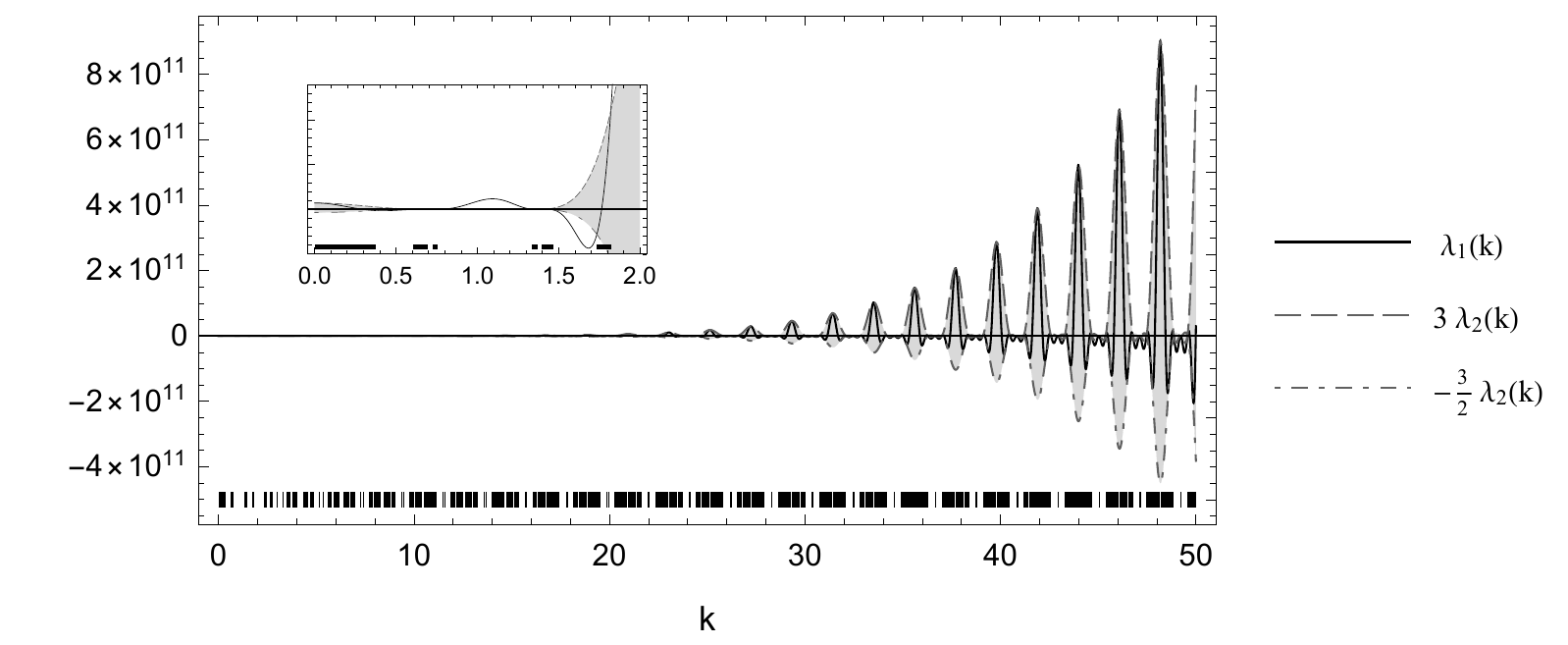}
  \label{KagEquilEx2}}
\caption{Spectral condition \eqref{Kag,gen,band,con} in the equilateral case for $c=1$ and $3$ with $\ell=1$. }
\label{KagEquilEx}
\end{figure}

These lattices can be regarded as a degenerate case of a kagome lattice when one of the edge lengths, $b$ or $c$, shrinks to zero. The elementary cell now contains a single vertex of degree six, cf. Fig.~\ref{KagTendTrig}.
 % -------------- %
 % -------------- %
Let us recall that the condition \eqref{Kag,Pos,SC,dc} is symmetric with respect to the interchange of $b=0$ and $c=0$. To find the spectral condition of the triangular lattice, we can either use the natural Ans\"atze and match them as we did when deriving \eqref{Kag,Pos,SC,dc}, or to take the limit $c\rightarrow d$ in the latter; this yields
 % -------------- %
\begin{align}\label{Trig,Pos,SC}
& \left(k^2 \ell ^2+1\right)\;\sin ^2 \textstyle{\frac{kd}2} \;\times \\
&\quad\Big( 3\big(k^4 \ell ^4+6 k^2 \ell ^2+1\big)+\big(3 k^4 \ell ^4+10 k^2 \ell ^2+3\big) (2 \cos kd +\cos 2kd )-4 \big(k^2 \ell ^2-1\big)^2 \cos ^2 \textstyle{\frac{kd}2}\,f_{\theta }\Big)=0.\nonumber\end{align}
 % -------------- %
These two methods are almost equivalent, with a small exception which we will mention in Sec.~\ref{TrigNeg,section} below. Let us look how the corresponding spectrum looks like considering again the positive and negative part separately.

\subsection{Positive Spectrum}
\label{TrigPos,section}
 % -------------- %
The positive spectrum of the triangular lattice consists again of two parts:
 % -------------- %

\paragraph*{\emph{i}. Infinitely degenerate eigenvalues}
\label{TrigPosFlat,section}
 % -------------- %
\begin{itemize}
\item For $k=\frac{2n\pi}{d}$ with $n\in\mathbb{N}$, the number $k^{2}$ belongs to the spectrum. Inspecting the expression in the `large' bracket in \eqref{Trig,Pos,SC} for $k=\frac{2n\pi}{d}$, we get
 % -------------- %
    \[ \frac{4}{d^4}\Big(  \left(d^4+16 \pi ^4 n^4 \ell ^4\right)(3-f_{\theta })+8 \pi ^2 d^2 n^2 \ell ^2 \left(f_{\theta }+6\right)  \Big)>0, \]
  % -------------- %
    which implies that the flat bands may not be embedded in the continuous spectrum.
\item As in the kagome lattice case, it may happen that some positive bands degenerate to a point; here this happens at $k=\ell^{-1}$ for $d=\ell\big((-1)^{n+1}+(6 n-3) \big)\frac{\pi}{6}$ with $n\in\mathbb{N}$.
\end{itemize}
 % -------------- %

\paragraph*{\emph{ii}. Continuous bands} \mbox{}\\
\label{TrigPosContinuous,section}
 % -------------- %
The rest of the spectrum has a band-gap structure determined by vanishing of the `large' bracket in \eqref{Trig,Pos,SC}. Note that the latter reduces to $\frac{8}{d^2} (2n-1)^2 \ell ^2\pi ^2>0$ for $k=(2n-1)\frac{\pi}{d}$, $n\in\mathbb{N}$; hence dividing the corresponding equation by
$\left(k^2 \ell ^2-1\right)^2 \cos ^2 \textstyle{\frac{kd}2}$, it can be brought into a more convenient form
 % -------------- %
\begin{equation}\label{TrigPosCond}
f_{\theta }=G(k):=\frac{2 k^2 \ell ^2 \sec ^2 \frac{kd}{2} +\left(3 k^4 \ell ^4+10 k^2 \ell ^2+3\right) \cos kd}{\left(k^2 \ell ^2-1\right)^2}.
\end{equation}
 % -------------- %
Consequently, a number $k^2$ belongs to the spectral bands if and only if $G(k)$ lies in the interval $[-\tfrac{3}{2},3]$. The band-gap pattern in dependence on $d$ is illustrated in Fig.~\ref{fig4}, and moreover, two examples for $d=1$ and $5$ are shown in Fig.~\ref{TrigEx}. We find that:
 % -------------- %
\begin{outline}
\1 The first positive band starts at zero if $ d\geq 2 \sqrt{3}\,\ell$, otherwise it remains separated from zero since $G(k)= 3+\frac{3}{2}(12\ell^2-d^2)k^2+\mathcal{O}(k^4)$ is greater than three for $ d< 2 \sqrt{3}\,\ell$ and small values of momentum $k$.
\1 In the high-energy regime, we have again two types of asymptotic behavior. To take a closer look at their structure, we rewrite the spectral condition \eqref{Trig,Pos,SC} in the form
 % -------------- %
\begin{equation}\label{Trig,Alter,Asy}
 \gamma_1(k)+\frac{\gamma_2(k)}{k^2}=\mathcal{O}(k^{-4}),\end{equation}
  % -------------- %
where
 % -------------- %
\begin{align}\label{Trig,Asy,Alpha}
&   \gamma_1(k)=4 \ell ^4 \,\cos ^2 \frac{kd}{2} \,\big(3 \cos kd-f_{\theta }\big)   ,\nonumber \\
&   \gamma_2(k)= 2 \ell ^2\big(10 \cos kd +5 \cos2kd+9+2(\cos kd+1)f_{\theta }\big)  .\nonumber
\end{align}
 % -------------- %
To satisfy the condition \eqref{Trig,Alter,Asy} for large values of $k$, the the leading term $\gamma_1(k)$ should be close to zero; this results in two types of spectral bands:
% -------------- %
\2 \textbf{Pairs of narrow bands in the vicinity of $k=(2n-1)\frac{\pi}{d}$, $n\in\mathbb{N}$} \\
The two bands around the points $k=(2n-1)\frac{\pi}{d}$ and the gap between them are of asymptotically constant width. To see that, we set $k=(2n-1)\frac{\pi}{d}+\delta$. Then we have $k^{-2}=\frac{d^2}{4n^2\pi^2}+\mathcal{O}(n^{-3})$ as $n\to\infty$; substituting these values into \eqref{Trig,Alter,Asy} and solving the resulting equation for $\delta$, we obtain
 % -------------- %
\[\delta=\frac{\sqrt{2}}{\pi  \ell  \sqrt{f_{\theta }+3}}\,\frac{1}{n}+\mathcal{O}(n^{-3}) .\]
 % -------------- %
Since the band edges correspond to $f_{\theta }=-\tfrac{3}{2}$ and $3$, the width of the bands and the gap between them are respectively equal to $\frac{4}{d\ell\sqrt{3}}+\mathcal{O}(n^{-1})$ and $\frac{8}{d\ell\sqrt{3}}+\mathcal{O}(n^{-1}) $ as $n\rightarrow \infty$.
% -------------- %
\2 \textbf{Pairs of wide bands in the vicinity of $k=2n\frac{\pi}{d}$, $n\in\mathbb{N}$} \\
These bands and the gaps between them grow asymptotically but not at the same rate; large values of $k$ belong to the spectrum if and only if
 % -------------- %
\begin{equation}\label{Trig,bandCondition,high}
-\frac{1}{2}\leq \cos kd\leq 1  ,\end{equation}
 % -------------- %
 with a relative error $\mathcal{O}(k^{-2})$. The function $\cos kd$ is periodic with the period $T=\tfrac{2\pi}{d}$, and solutions of the equation $\cos kd=-\tfrac{1}{2}$ in the period are $\tfrac{2\pi}{3d}$ and $\tfrac{4\pi}{3d}$. On the other hand, $\cos kd$ for $k=\tfrac{\pi}{d}$ is $-1$; and $\cos kd$ is less than $-\tfrac{1}{2}$ over the domain $\left(\tfrac{2\pi}{3d},\tfrac{4\pi}{3d}\right)$. Consequently, the probability of belonging to the spectrum is equal to
 % -------------- %
\[ P_{\sigma}(H)= 1-\frac{1}{T}\left( \frac{4\pi}{3d}-\frac{2\pi}{3d} \right)=\frac{2}{3}.  \]
 % -------------- %
 The triangular lattice is equilateral, hence it makes no sense to speak about the universality. It is nevertheless interesting that the above value coincides with \eqref{probequi}. We note also that these bands appear in pairs centered around the points $k=2n\frac{\pi}{d}$ marking the flat bands; as mentioned earlier, they may not be embedded in the continuous spectrum; from \eqref{Trig,bandCondition,high} we see that these gaps are situated in the vicinity of the Brillouin zone center.
 % -------------- %
\1 As in the kagome case, the on-shell scattering matrix is nontrivial at high energies. Now the vertex degree is six and from \eqref{sij,onshell} we get
 % -------------- %
 \[ \underset{k\to \infty }{\text{lim}}S(k)=\frac{2}{3}\, I_6\,+ \frac13 {\scriptsize
 \left(
\begin{array}{cccccc}
 \phantom{-}0 & \phantom{-}1 & -1 & \phantom{-}1 & -1 & \phantom{-}1 \\
 \phantom{-}1 & \phantom{-}0 & \phantom{-}1 & -1 & \phantom{-}1 & -1 \\
-1 & \phantom{-}1 & \phantom{-}0 & \phantom{-}1 & -1 & \phantom{-}1 \\
 \phantom{-}1 & -1 & \phantom{-}1 & \phantom{-}0 & \phantom{-}1 & -1 \\
 -1 & \phantom{-}1 & -1 & \phantom{-}1 & \phantom{-}0 & \phantom{-}1 \\
\phantom{-}1 & -1 & \phantom{-}1 & -1 & \phantom{-}1 & \phantom{-}0 \\
\end{array}
\right) }  .\]
% -------------- %
\1 As $d\rightarrow\infty$, the number of bands in a fixed interval increases, roughly linearly with $d$, while the probability to be in the spectrum is asymptotically constant.
\end{outline}
 % -------------- %

\subsection{Negative Spectrum}
\label{TrigNeg,section}
\begin{figure}[!htb]
\centering
\includegraphics[scale=1.2]{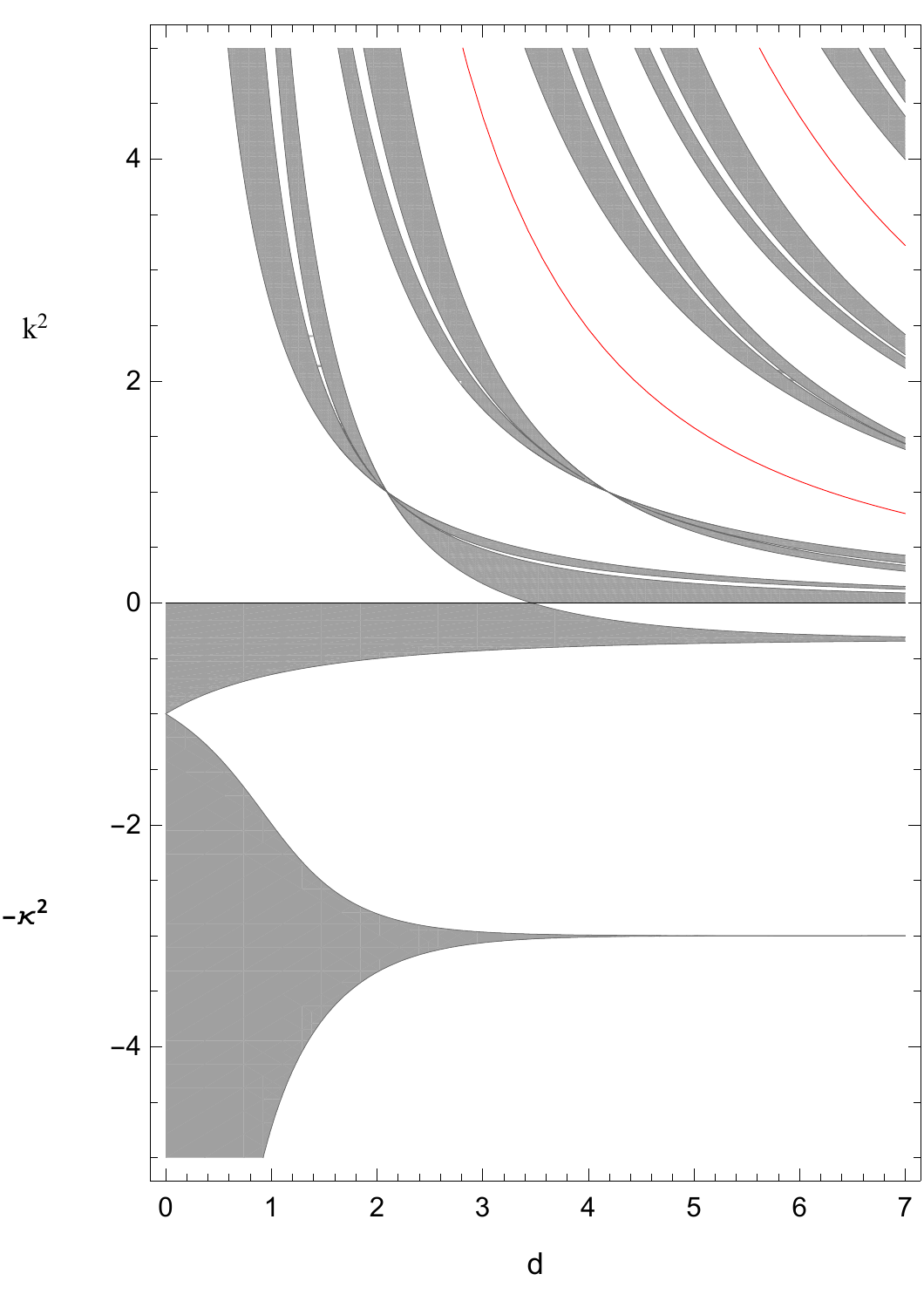}
\caption{ Spectrum of the triangular lattice in dependence on $d$ for $\ell=1$. The red curves correspond to the flat bands $\frac{4n^2\pi^2}{d^2}$ for $n=1,2$. On the other hand, the points $(2\frac{\pi}{3},1)$ and $(4\frac{\pi}{3},1)$ correspond to the flat bands of the second bullet point in Sec.~\ref{TrigPos,section}.}
\label{fig4}
\end{figure}

\begin{figure}[!htb]
  \centering
  \subfloat[Spectral bands for $d=1$; the first band remains separated from zero.]{\includegraphics[width=0.75\textwidth]{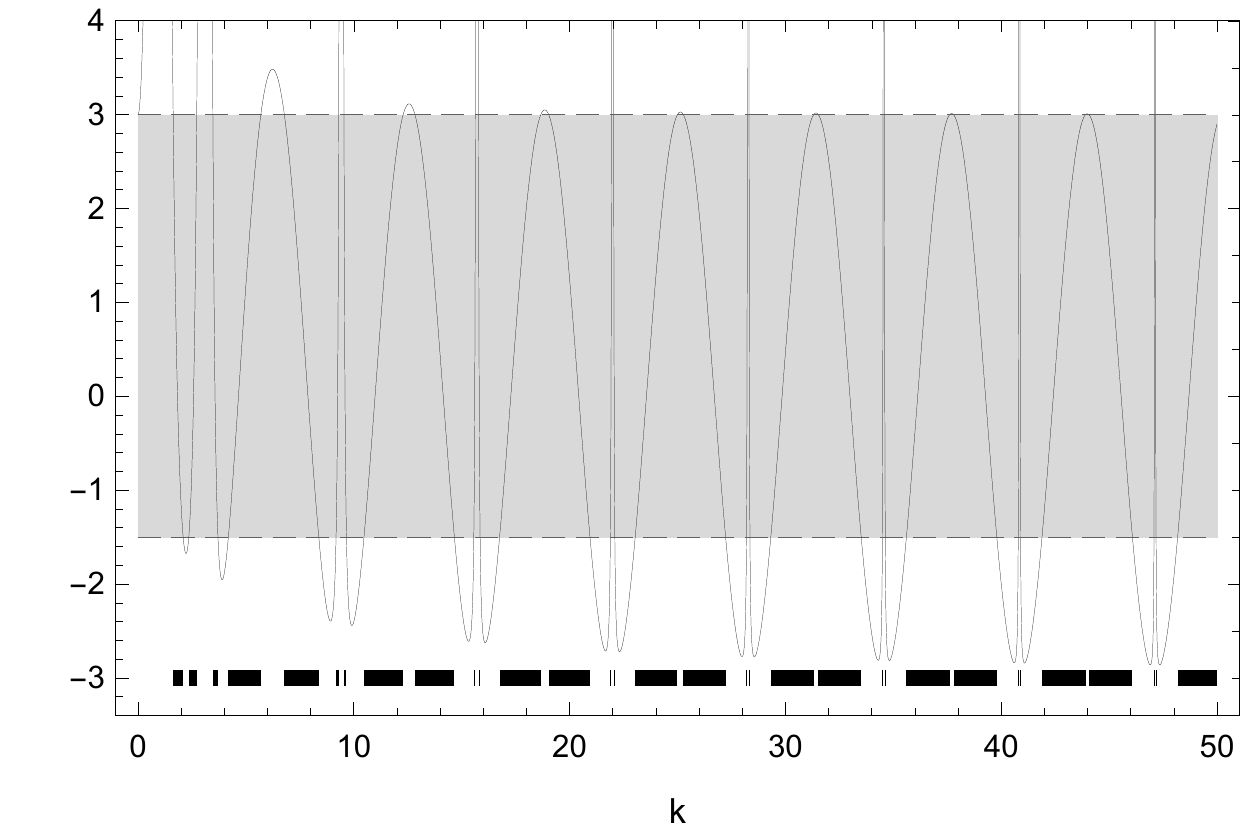}\label{TrigEx1}}
  \hfill
  \subfloat[Spectral bands for $d=5$; the first band starts at zero.]{\includegraphics[width=0.75\textwidth]{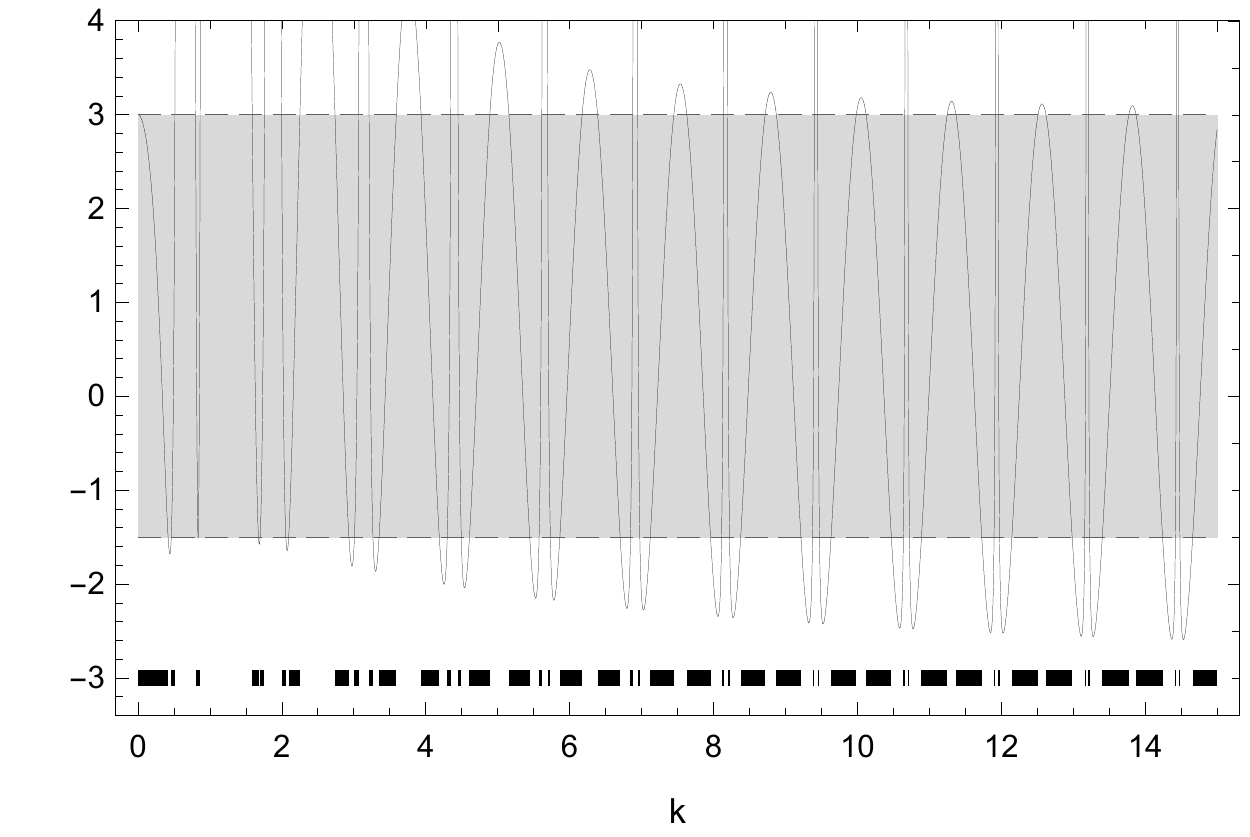}\label{TrigEx2}}
\caption{Spectral condition \eqref{TrigPosCond} for $d=1$ and $5$ with $\ell=1$. }
\label{TrigEx}
\end{figure}

\begin{figure}[!htb]
\centering
\includegraphics[scale=1.2]{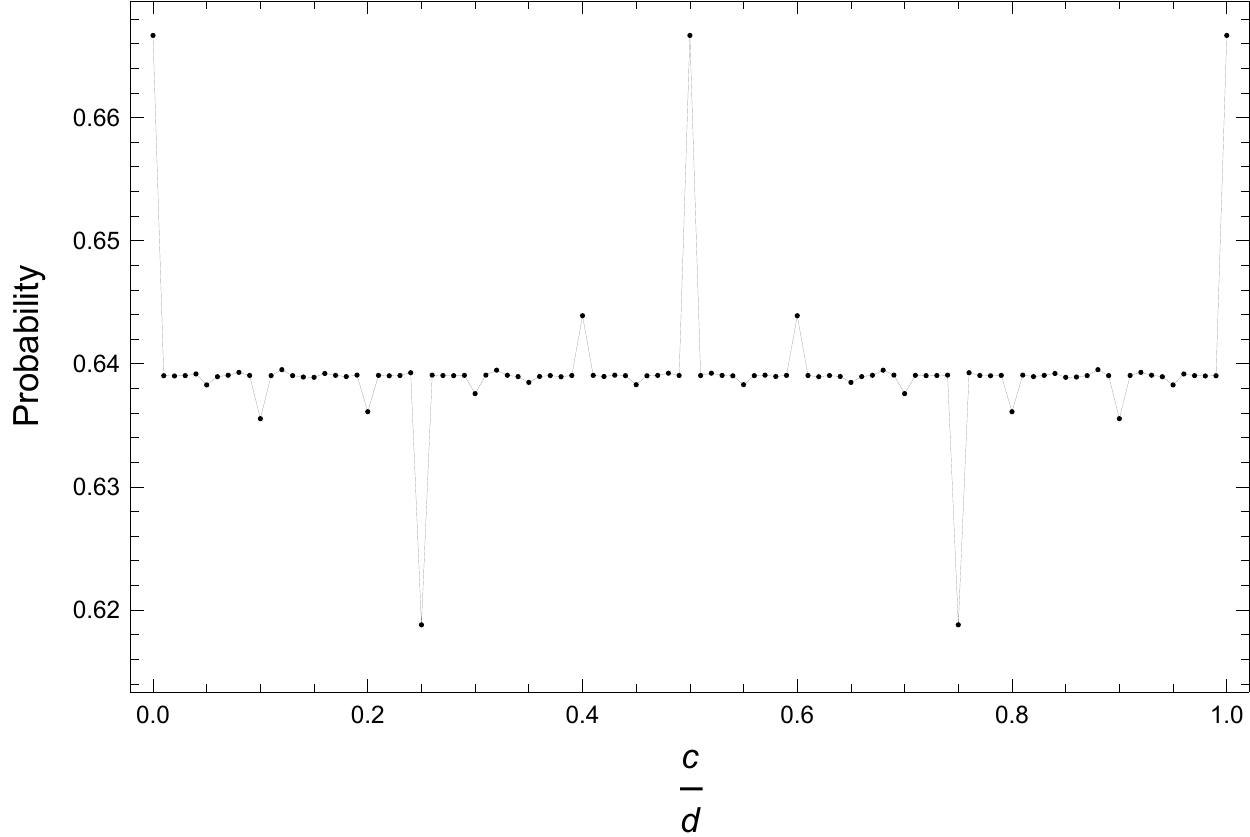}
\caption{The probability \eqref{probsigma} as a function of the edge length ratio.}
\label{probfig}
\end{figure}

\begin{figure}[!htb]
\centering
\includegraphics[scale=.8]{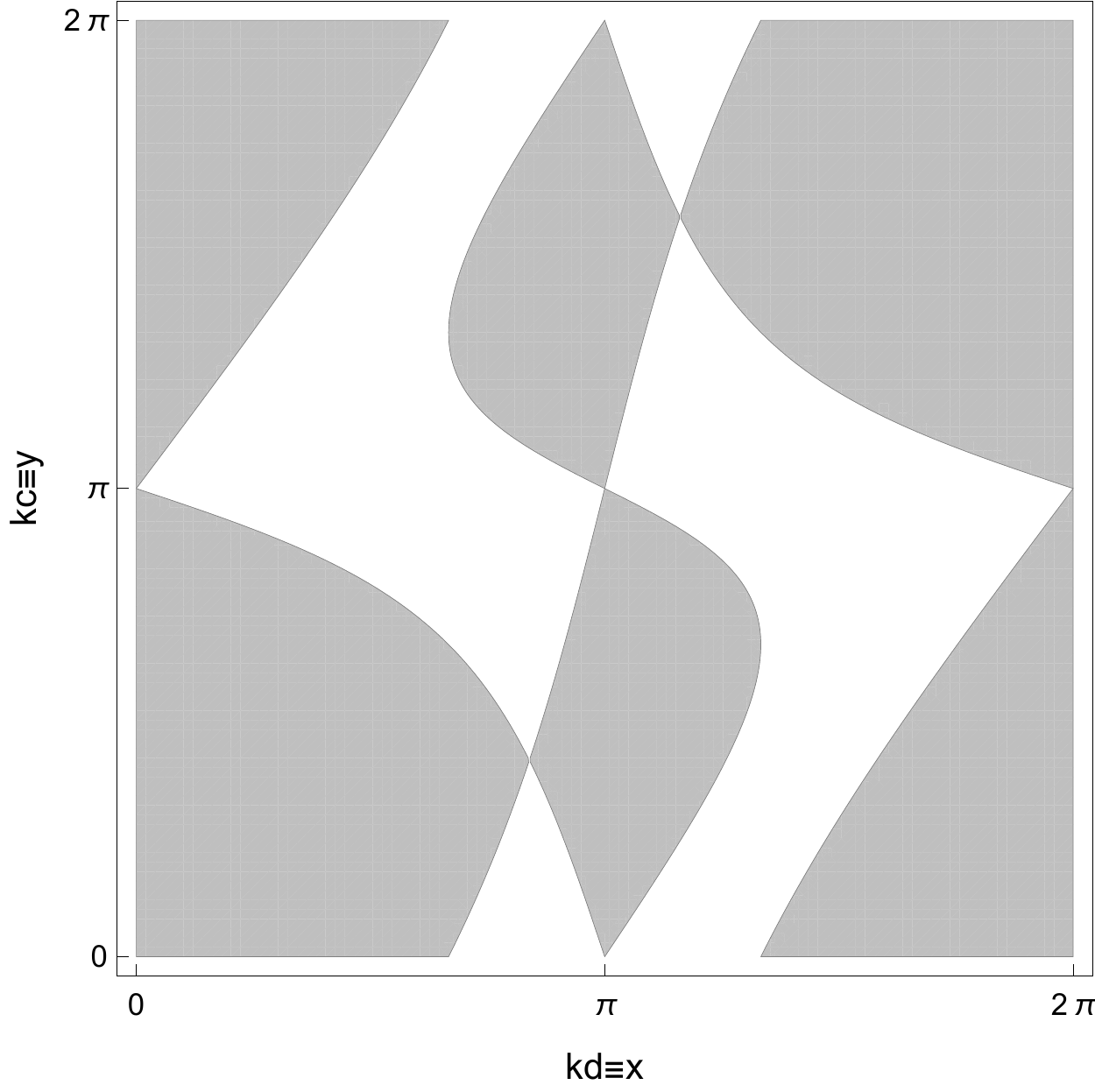}
\caption{The probability \eqref{probsigma} calculated as the area of the indicated region.}
\label{probfig2}
\end{figure}

 % -------------- %
To find the negative spectrum, we may replace $k$ by $i\kappa$ in \eqref{Trig,Pos,SC}, which leads to the condition
 % -------------- %
 \begin{align}\label{TrigNegSC}
 & \left(\kappa ^2 \ell ^2-1\right)\; \sinh ^2 \frac{\kappa d}{2}  \; \times \\
 & \left(3 \big(\kappa ^4 \ell ^4-6 \kappa ^2 \ell ^2+1\big)+\big(3 \kappa ^4 \ell ^4-10 \kappa ^2 \ell ^2+3\big)\big(2 \cosh\kappa d+\cosh 2\kappa d \big)-4 \big(\kappa ^2 \ell ^2+1\big)^2 \cosh ^2 \frac{\kappa d }{2} \; f_{\theta }\right)=0.\nonumber
 \end{align}
  % -------------- %
Let us mention first the exception mentioned in the opening of the section. It may seem that the energy $-\ell^{-2}$ belongs to the spectrum but in reality it is a spurious solution. To see that, we note that for the triangular lattice the functions $\chi_{2},\,\chi_{3},\,\varphi_2,\, \varphi_3,\,\psi_3$, and $\psi_4$ in \eqref{KagAnsatz} are absent; rewriting then the matching condition \eqref{KagEqs} accordingly, and computing the determinant of the corresponding system at $k=i \,\ell^{-1}$, we arrive at the expression
 % -------------- %
\[ 1024 \,i \,\e^{2 i \theta _2}\,\ell ^{-3}\,\sinh ^2 \frac{d}{2 \ell } \,\left( 3+2\,f_\theta+2 \left(f_\theta+1\right)\cosh \frac{d}{\ell }+\cosh \frac{2d}{\ell }  \right),\]
  % -------------- %
which is nonzero taking into account the range of $f_\theta$ and the fact that $\cosh 2x>\cosh x$ holds for $x>0$, and consequently, the point $-\ell^{-2}$ cannot belong to the spectrum. The true spectral condition comes from vanishing of the `large' bracket in \eqref{TrigNegSC}; we can rewrite it in the form
 % -------------- %
\begin{equation}\label{TrigNegCond}
f_{\theta }=\tilde{G}(\kappa):=\frac{\left(3 \kappa ^4 \ell ^4-10 \kappa ^2 \ell ^2+3\right) \cosh\kappa d -2 \kappa ^2 \ell ^2 \text{sech}^2 \frac{\kappa d }{2} }{\left(\kappa ^2 \ell ^2+1\right)^2}.
\end{equation}
 % -------------- %
In other words, a number $-\kappa^2$ belongs to the spectrum if $\tilde{G}(\kappa)$ lies in the interval $[-\tfrac{3}{2},3]$. The negative spectrum in dependence on the length $d$ is illustrated in Fig.~\ref{fig4}. We see that:
 % -------------- %
\begin{itemize}
\item The spectrum consists of two bands. We first note that $\tilde{G}(\ell^{-1})=-\cosh\frac{d}{\ell }-(\cosh\frac{d}{\ell }+1)^{-1}<-\tfrac{3}{2} \,$ which can be easily checked by computing the derivative of the right-hand side with respect to~$d$, equal to $\tfrac{1}{\ell}\sinh\tfrac{d}{\ell}\left( ( \cosh\tfrac{d}{\ell}+1 )^{-2}-1 \right)<0$. Hence $d\mapsto \tilde{G}(\ell^{-1})$ is decreasing reaching the value $-\frac32$ at $d=0$. On the other hand, we have $\tilde{G}(0)=3$ and $\lim_{\kappa\to\infty} \tilde{G}(\kappa)=+\infty$. Consequently, there is at least one negative band in each of the domains $(0,\ell^{-1})$ and $(\ell^{-1},\infty)$, however, according to Theorem 2.6 in \cite{BET21} the lattice cannot have more than two negative bands, since the matrix $U$ describing the coupling \eqref{genbc} in a vertex of degree six has by \eqref{Neg,Eig,StarG} exactly two eigenvalues in the upper complex halfplane.
% -------------- %
\item For $ d\leq 2 \sqrt{3}\,\ell$, the first negative reaches zero, while for $ d> 2 \sqrt{3}\,\ell$, the spectrum remains separated from zero since $ \tilde{G}(\kappa)=3+\textstyle{\frac{3}{2}}(d^2-12\ell^2)\kappa^2+\mathcal{O}(\kappa^4)$ is greater than three for $ d>2\sqrt{3}\,\ell$ and small values of $\kappa$.
% -------------- %
\item For large values of $d$, the negative bends become exponentially narrow and approach the eigenvalues of a star graph of degree six. Since we have $\cosh\kappa d \approx \frac{1}{2} \e^{\kappa d}$ as $d\rightarrow\infty$, one can write the spectral condition \eqref{TrigNegSC} as
% -------------- %
    \[   \tilde{f}(\ell ;\kappa )\,\e^{2\kappa d }+ \tilde{g}(\ell ;\kappa ,f_{\theta })\,\e^{\kappa d }+\tilde{h}(\ell ;\kappa ,f_{\theta })+\mathcal{O}(\e^{-\kappa d })=0      ,\]
% -------------- %
    with
% -------------- %
\begin{align}
& \tilde{f}(\ell ;\kappa ):= \textstyle{\frac{1}{2}} \big( 3 \kappa ^4 \ell ^4-10 \kappa ^2 \ell ^2+3 \big)    ,   \nonumber\\
& \tilde{g}(\ell ;\kappa ,f_{\theta }):=\big( 3 \kappa ^4 \ell ^4-10 \kappa ^2 \ell ^2+3 \big)- \left(\kappa ^2 \ell ^2+1\right)^2 \,f_{\theta }    ,   \nonumber \\
& \tilde{h}(\ell ;\kappa ,f_{\theta }):= 3 \left(\kappa ^4 \ell ^4-6 \kappa ^2 \ell ^2+1\right)-2 \left(\kappa ^2 \ell ^2+1\right)^2 \,f_{\theta } .   \nonumber \end{align}
% -------------- %
In the limit $d\rightarrow\infty$, the bands therefore shrink to the points determined by the condition $\tilde{f}(\ell ;\kappa )=0$, or explicitly to the energies $-3\ell^{-2}$ and $-\frac{1}{3}\ell^{-2}$ which fits with \eqref{Neg,Eig,StarG}. To estimate the widths of the shrinking bands, we set $\kappa_1=\sqrt{3}\ell^{-1}+\delta_1$ and $\kappa_2=(\ell\sqrt{3})^{-1}+\delta_2$; solving then the resulting equations for $\delta_1$ and $\delta_2$, we obtain the following asymptotic expressions for the energy and width of the bands
 % -------------- %
\begin{align}
  E_{1}=-\kappa_{1}^{2}(f_{\theta })&= -3\ell^{-2}-4\ell^{-2}\,\e^{-\frac{\sqrt{3} d}{\ell }}\,f_{\theta }+\mathcal{O}(\e^{-\frac{2 \sqrt{3} d}{\ell }})           ,   \nonumber \\
 \Delta E_{1}&= 18\ell^{-2}\e^{-\frac{\sqrt{3} d}{\ell }}+\mathcal{O}(\e^{-\frac{2 \sqrt{3} d}{\ell }})             ,   \nonumber \end{align}
 % -------------- %
\begin{align}
  E_{2}=-\kappa_{2}^{2}(f_{\theta })&= -\textstyle{\frac{1}{3}}\ell^{-2}     +\textstyle{\frac{4}{9}}\ell^{-2}\,\e^{-\frac{d}{\sqrt{3} \ell }} \,f_{\theta }+\mathcal{O}(\e^{-\frac{2 d}{\sqrt{3} \ell }})             ,   \nonumber \\
  \Delta E_{2}&= 2 \ell^{-2} \e^{-\frac{d}{\sqrt{3} \ell }}+\mathcal{O}(\e^{-\frac{2 d}{\sqrt{3} \ell }})              .   \nonumber \end{align}
 % -------------- %
\end{itemize}

 % -------------- %
%\pagebreak
 % -------------- %

 % -------------- %
%\pagebreak
 % -------------- %

 % -------------- %
%\pagebreak
 % -------------- %

 % -------------- %
%\pagebreak
 % -------------- %

 % -------------- %
%\pagebreak
 % -------------- %

 % -------------- %
%\pagebreak
 % -------------- %

 % -------------- %
%\pagebreak
 % -------------- %

 % -------------- %

 % -------------- %
%\pagebreak
 % -------------- %

 % -------------- %
%\pagebreak
 % -------------- %

 % -------------- %
%\pagebreak
 % -------------- %

\subsection*{Acknowledgments}
The research was supported by the Czech Science Foundation project 21-07129S and by the EU project CZ.02.1.01/0.0/0.0/16\textunderscore 019/0000778. MB's work was also supported by the Internal Postdoc Project UHK for years 2021-2022.

\pagebreak
 % -------------- %

\end{document}